\documentclass[pre, twocolumn, superscriptaddress, floatfix, nofootinbib]{revtex4-1}  

\usepackage[paperwidth=210mm,paperheight=297mm,centering,hmargin=2cm,vmargin=2.5cm]{geometry}

\usepackage[pdftex]{graphicx}
\usepackage{amsmath,amsfonts,amssymb}
\usepackage{MnSymbol}
\usepackage{dsfont}
\usepackage{enumitem}
\usepackage{csquotes}

\usepackage{hyperref}
\usepackage{xcolor}

\definecolor{bluemoi}{rgb}{0.25,0.50 ,0.75} 

\hypersetup{
  bookmarksopen=true,
  pdftitle=" ",
  pdfauthor=" ", 
  pdfsubject=" ", 
  pdftoolbar=true, 
  pdfmenubar=true, 
  pdfhighlight=/O, 
  colorlinks=true, 
  pdfpagemode=UseNone, 
  pdfpagelayout=SinglePage, 
  pdffitwindow=true, 
  linkcolor=bluemoi, 
  citecolor=bluemoi, 
  urlcolor=bluemoi 
}

\makeatletter
\renewcommand\@biblabel[1]{#1} 

\renewcommand\newblock{\hskip .11em\@plus.33em\@minus.07em}
\makeatother

\setcitestyle{authoryear,round}
\bibpunct{(}{)}{;}{a}{,}{}

\makeatletter
\newcommand{\removeperiod}{\@ifnextchar.{\@gobble}\relax}
\makeatother 

\makeatletter
\renewcommand{\figurename}{\sf \textbf{Figure}}
\renewcommand{\thefigure}{\arabic{figure}}
\renewcommand{\fnum@figure}{\sf\textbf{\figurename}~\textbf{\thefigure}}
\renewcommand{\tablename}{\sf\textbf{Table}}
\renewcommand{\thetable}{\arabic{table}}
\renewcommand{\fnum@table}{\sf\textbf{\tablename}~\textbf{\thetable}}
\makeatother

\begin{document}

\title{Biogeographical network analysis of plant species distribution\\ in the Mediterranean region} 

\author{Maxime Lenormand}
\thanks{Corresponding authors: maxime.lenormand@irstea.fr \& guillaume.papuga@gmail.com who contributed equally to this work.}
\affiliation{Irstea, UMR TETIS, 500 rue JF Breton, 34093 Montpellier, France}

\author{Guillaume Papuga}
\thanks{Corresponding authors: maxime.lenormand@irstea.fr \& guillaume.papuga@gmail.com who contributed equally to this work.}
\affiliation{Conservatoire botanique national m{\'e}diterran{\'e}en de Porquerolles, Parc scientifique Agropolis, 2214 boulevard de la Lironde, 34980 Montferrier sur Lez, France}
\affiliation{UMR 5175 CEFE, CNRS, 1919 route de Mende, 34293 Montpellier cedex 5, France}

\author{Olivier Argagnon}
\affiliation{Conservatoire botanique national m{\'e}diterran{\'e}en de Porquerolles, Parc scientifique Agropolis, 2214 boulevard de la Lironde, 34980 Montferrier sur Lez, France}

\author{Maxence Soubeyrand}
\affiliation{Irstea, UMR TETIS, 500 rue JF Breton, 34093 Montpellier, France}

\author{Guilhem De Barros}
\affiliation{Conservatoire botanique national m{\'e}diterran{\'e}en de Porquerolles, Parc scientifique Agropolis, 2214 boulevard de la Lironde, 34980 Montferrier sur Lez, France}

\author{Samuel Alleaume}
\affiliation{Irstea, UMR TETIS, 500 rue JF Breton, 34093 Montpellier, France}

\author{Sandra Luque}
\affiliation{Irstea, UMR TETIS, 500 rue JF Breton, 34093 Montpellier, France}

\begin{abstract}
\vspace*{0.5cm}	
The delimitation of bioregions helps to understand historical and ecological drivers of species distribution. In this work, we performed a network analysis of the spatial distribution patterns of plants in south of France (Languedoc-Roussillon and Provence-Alpes-C\^ote d'Azur) to analyze the biogeographical structure of the French Mediterranean flora at different scales. We used a network approach to identify and characterize biogeographical regions, based on a large database containing 2.5 million of geolocalized plant records corresponding to more than 3,500 plant species. This methodology is performed following five steps, from the biogeographical bipartite network construction, to the identification of biogeographical regions under the form of spatial network communities, the analysis of their interactions and the identification of clusters of plant species based on the species contribution to the biogeographical regions. First, we identified two sub-networks that distinguish Mediterranean and temperate biota. Then, we separated eight statistically significant bioregions that present a complex spatial structure. Some of them are spatially well delimited, and match with particular geological entities. On the other hand fuzzy transitions arise between adjacent bioregions that share a common geological setting, but are spread along a climatic gradient. The proposed network approach illustrates the biogeographical structure of the flora in southern France, and provides precise insights into the relationships between bioregions. This approach sheds light on ecological drivers shaping the distribution of Mediterranean biota: the interplay between a climatic gradient and geological substrate shapes biodiversity patterns. Finally this work exemplifies why fragmented distributions are common in the Mediterranean region, isolating groups of species that share a similar eco-evolutionary history.
\end{abstract}

\maketitle

\section*{Introduction}

The delimitation of biogeographical regions or bioregions based on the analysis of their biota has been a founding theme in biogeography, from the pioneer work of  \citet{Wallace1876}, \citet{Murray1866} or \citet{Wahlenberg1812} to the most recent advances of \citet{Cheruvelil2017, Ficetola2017}. Describing spatial patterns of biodiversity has appeared fundamental to understand the historical diversification of biota, and gain a better understanding of ecological factors that imprint spatial patterns of biodiversity \citep{Ricklefs2004, Graham2006}. Additionally, it has become a key element in the identification of spatial conservation strategies \citep{Funk2002, Rushton2004, Mikolajczak2015}. To divide a given territory into meaningful and coherent bioregions, the overall aim is to minimize the heterogeneity in taxonomic composition within regions, while maximizing differences between them \citep{Stoddart1992,Kreft2010}. Although such delineation of bioregions has been based for a long time on expert knowledge of qualitative data collection the increasing availability of species-level distribution data and recent technological advances have allowed for the development of more rigorous frameworks \citep{Kreft2010}. Multivariate methods, such as hierarchical clustering algorithms, have thus been successfully applied in a wide range of studies focused on a variety of organisms, under very different spatial scale (from regional to worldwide perspective). Yet, the production of detailed cartographic outputs portraying the differentiation of vegetation into distinct homogeneous bioregions remains difficult, especially where spatial heterogeneity of assemblages is associated with complex environmental gradients \citep{Mikolajczak2015}. Besides, the identification of meaningful and coherent bioregions represents only one step of the biogeographical regionalizations \citep{Morrone2018}. It is also crucial to propose new metrics to quantify the relationship between bioregions and to analyze species and spatial relationships.

Some regions of the world oppose inherent difficulties due to their highly diversified biota, reflecting complex eco-evolutionary processes. The Mediterranean basin is one of the largest and most important biodiversity hotspots in the world \citep{Myers2000,Blondel2010}. This region hosts about 25,000 plant species representing 10\% of the world's total floristic richness concentrated on only 1\% of the world's surface \citep{Greuter1991}. Additionally, a high level of narrow endemism is a major feature of this biome \citep{Thompson2005a}. Endemism and richness result in a very heterogeneous region, whose comprehension of spatial patterns of plant distribution is clue to get better insights into past and actual processes shaping biodiversity \citep{Quezel1999}. The onset of the Mediterranean climate during the Pliocene and the diverse glacial periods of the Pleistocene \citep{Quezel2004} have shaped the most important phases of plant evolution since the Tertiary \citep{Thompson2005a}. Additionally, due to a long history of human presence, contemporary flora has been widely influenced by human-mediated dispersal, land-use and other pressures \citep{Dahlin2014,Fenu2014}. The French Mediterranean area stretches from the Pyrenees in the south-west to the slopes of the Maritime and Ligurian Alps in the east. It encompasses three zones highlighted as glacial refugia \citep{Medail2009}, and the eastern sector represents one of the ten main biodiversity hotspots in the Mediterranean area \citep{Medail1997}. This area represents the northern limit of the Mediterranean climate in the western basin, and thus constitutes a climatic transition from a Mediterranean zone that has a summer drought to a temperate zone less prone to summer drought \citep{Walter1991,Walter1994}. On a finer scale, the climate is more complex with several subtypes and intricated boundaries \citep{Joly2010, Tassin2017}. Several works have tried to map the distribution of biogeographical entities. To date, no statistical analysis had been ran to tackle those expert-based maps with up-to-date plant records, in order to test their reliability.

In order to depict spatial structure in such a complex regional flora, a large dataset is required. While the level of diversity and complexity of such dataset may appear overwhelming at first glance, the emergence of network-based approaches has opened new paths for identifying and delimiting bioregions where the presence-absence matrix is represented by a bipartite network. For example, \citet{Kougioumoutzis2014} applied the NetCarto algorithm \citep{Guimera2005} in order to identify biogeographical modules within the phytogeographical area of the Cyclades. Similarly, \citet{Vilhena2015} proposed a network approach for delimiting biogeographical region based on the InfoMap algorithm \citep{Rosvall2008}. By embedding species distributional data into complex networks, these methods have the great advantage to be generic, flexible and to incorporate several scales in the analysis. Most importantly, these methods integrate species community and spatial units within a single framework, which allow to test the relative contribution of each taxa to bioregions depicted, and to represent the relationship between those bioregions based on those contributions. 

In this study, we present a biogeographical network analysis of plant species distribution in the French Mediterranean area at different scales. The French Mediterranean territory represents an interesting study area to test new approaches, given the excellent knowledge of the spatial distribution of the plant species revealed by botanical inventories \citep{Tison2014, Tison2014a} and the detailed databases compiled by the French National Botanic Conservatory of Porquerolles and the Alpine National Botanic Conservatory.  The objective of this work is to delineate bioregions, identify groups of species and analyse the relationships between the two entities.

\begin{figure*}
	\begin{center}
		\includegraphics[width=16cm]{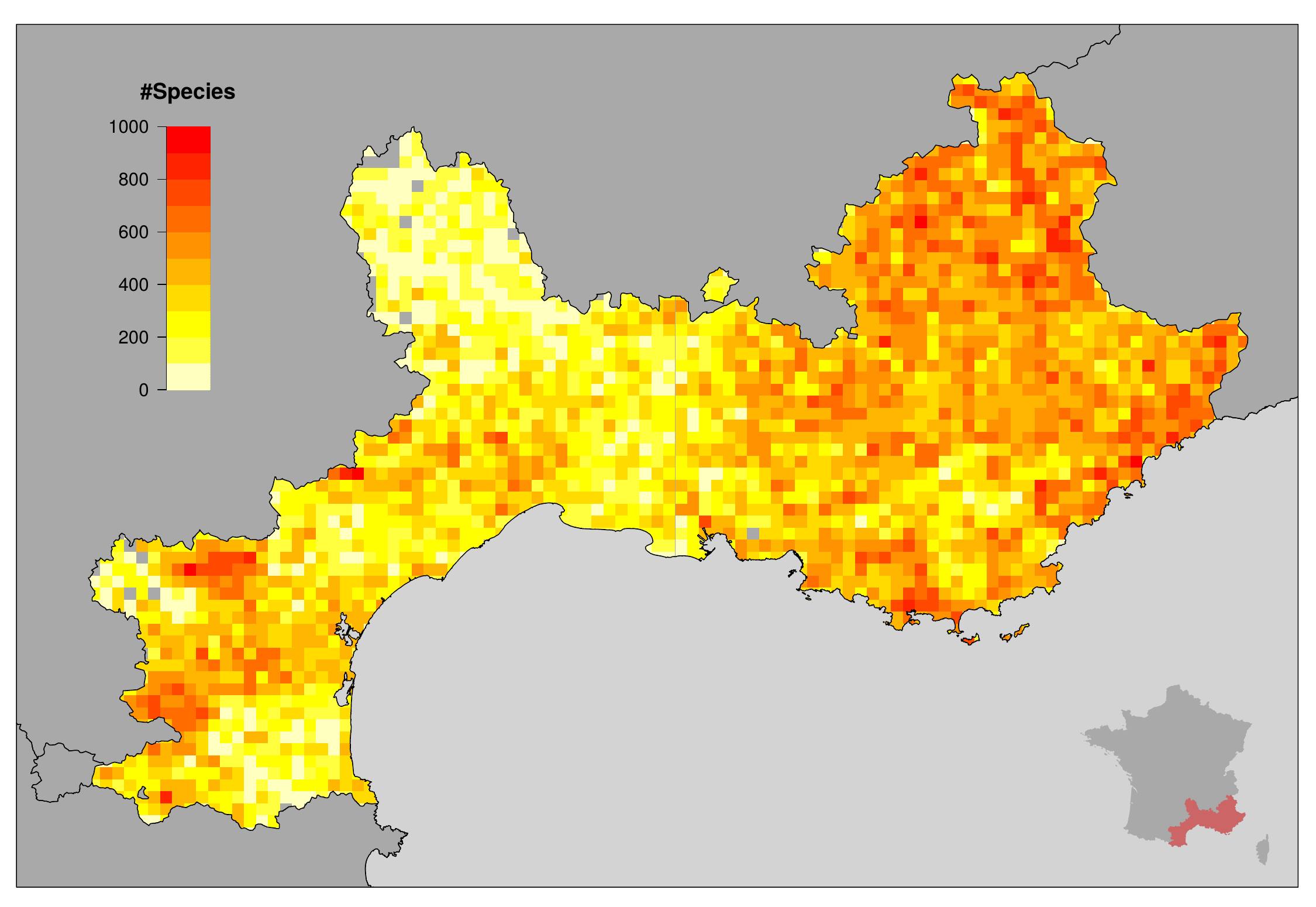}
		\caption{\sf \textbf{Distribution of the number of species per grid cell (l = 5 km).} The inset shows a map of France including the studied area colored in red. An altitude map of the studied area is available in Appendix. \label{Fig1}}
	\end{center}
\end{figure*} 

\section*{Materials and Methods}

\subsection*{Dataset and study area}

The study area, situated in southern France, encompasses the former Languedoc-Roussillon region (five departments of the current Occitanie region: Pyr{\'e}n{\'e}es-Orientales, Aude, H{\'e}rault, Gard and Loz{\`e}re) and the whole Provence-Alpes-C{\^o}te d'Azur region. It extends around the entire Mediterranean coastline of mainland France and inland, comprising almost all the Mediterranean hinterland, totalling $558,776 \mbox{ km}^2$ (Figure \ref{Fig1}). The topography is structured by three major mountain ranges, the Pyrenees in the southwest, the Massif central in the north-west and the Maritime Alps in the north-east. In-between, the landscape is mostly hilly with some lowlands around rivers that flow into lagoons or marshy deltas such as the Camargue. The Rh{\^o}ne is the main structuring river and delimitates western and eastern subregions. Acidic substrates and silicate soils are mainly found in the aforementioned mountain ranges and in the smaller Maures-Est{\'e}rel range in southern Provence. The remaining part of the territory is dominated by calcareous or marly substrates (principally Cretaceous and Jurassic), with some significant alluvial zones and small volcanic areas. 

The SILENE database\footnote[1]{Conservatoire Botanique National M{\'e}diterran{\'e}en \& Conservatoire Botanique National Alpin (Admin.). AAAA. SILENE-Flore [online]. \url{http://flore.silene.eu} (accessed the 16/03/2018)}, has been created in 2006, and is the reference botanical database in the study area. It contains historical data gathered from the scientific literature and herbaria along with more recent data coming from public studies, partnerships, local amateur botanist networks and professional botanists of the Botanical Conservatory. Our analysis is based on a $5 \times 5$ km$^2$ grid cells. We decided to only retain data whose georeferencement precision is below 10 meters. While the SILENE database contained nearly five million observations at the date of the export (June 2016), we deleted several taxa whose distribution is still insufficiently known and could distort the results (e.g. apomictic taxa such as \textit{Rubus} or \textit{Hieracium}). For the same reason, we also aggregated all sub-taxa at the species level. The final dataset results in $4,263,734$ vegetation plant samples corresponding to $3,697$ plant species. We divided the study area using a UTM grid composed of $2,607$ squares of lateral size $l=5$ km. In order to assess the impact of the spatial resolution on the results \citep{Lennon2001,Divisek2016}, we also applied the aforementioned biogeographical network analysis with a grid composed of squares of lateral size $l=10$ km (see Figure S\ref{FigS1} and Table S\ref{TabS1} in Appendix for more details).

\begin{figure*}
	\begin{center}
		\includegraphics[width=16cm]{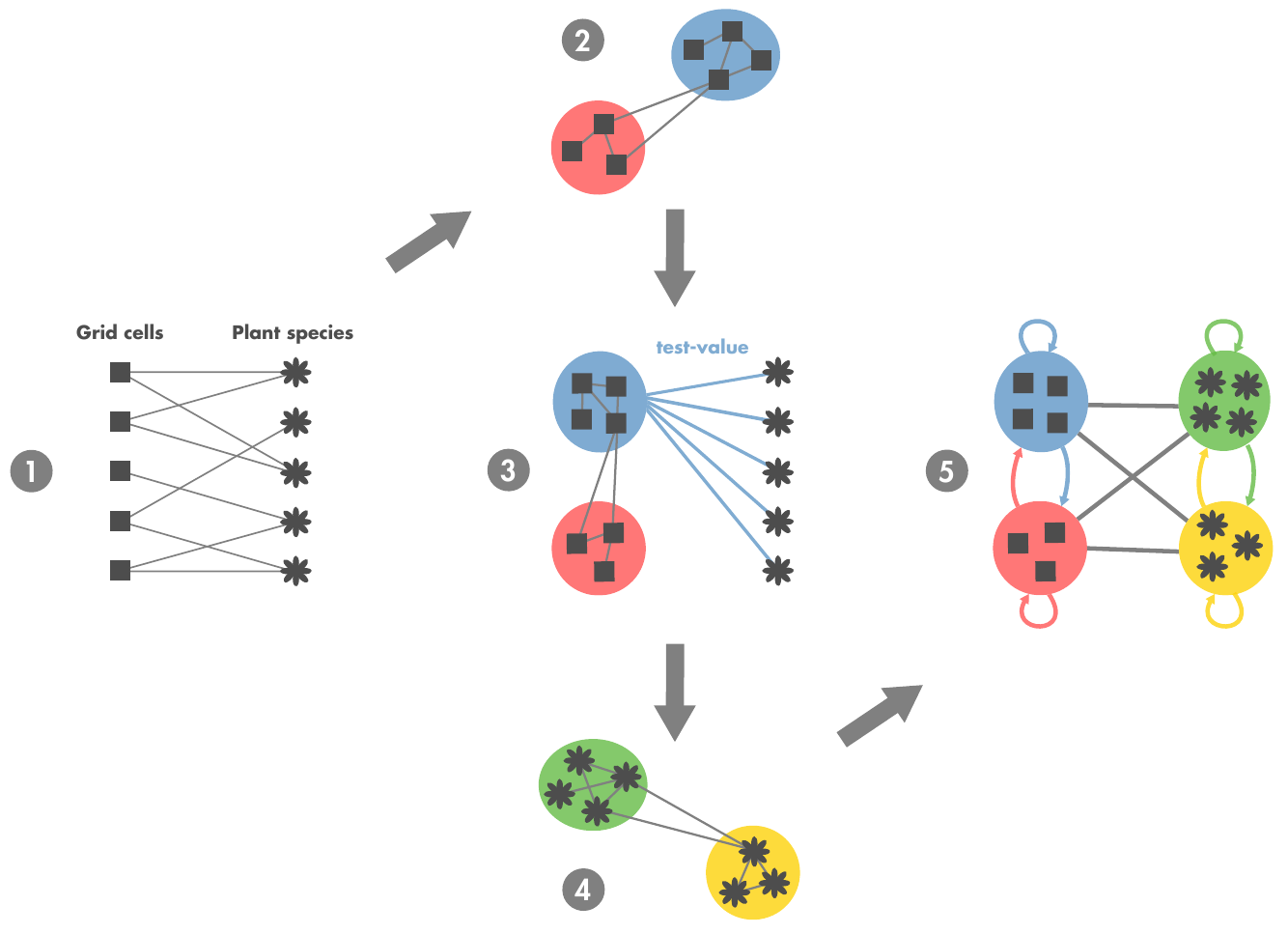}
		\caption{\sf \textbf{Steps of the biogeographical network analysis.} 1. Biogeographical bipartite network where grid cells and species are linked by the presence of a species (or a group of species) in a given grid cell during a certain time window. Note that there is no link between nodes belonging to the same set. 2. The bipartite network is then spatially projected by using a similarity measure of species composition between grid cells. Bioregions are then identified with a network community detection algorithm. 3. The test-value matrix based on the contribution of species to bioregions is computed. 4. Then, a network of similarity between species is built, based on the test-value matrix. Groups of species sharing similar spatial features are identified using a community detection algorithm. 5. Finally, a coarse-grained biogeographical network unveiling the biogeographical structure of the studied area and the relationship between bioregions is obtained. \label{Fig2}}
	\end{center}
\end{figure*}

\subsection*{Biogeographical network analysis}

\noindent \textbf{1. Biogeographical bipartite network.} Delineating bioregions requires a link between the species studied and their spatial environment. This link is usually identified with presence-absence matrices where each row represents a grid cell and each column a species. The region of interest is usually divided into grid cells, the resolution of which depends mostly on the size of the study area, the taxonomic group under study and the accuracy of the data. According to the type and quality of data, but also to the research question, the species \emph{relev\'e} can be aggregated both spatially or by group of species. Another way of formalizing complex interactions between species and grid cells is to build a biogeographical bipartite network. This bipartite network enables us to model relations between two disjoint sets of nodes, grid cells and species (in our case), which are linked by the presence of a species (or a group of species) in a given grid cell during a certain time window (Step 1 in Figure \ref{Fig2}). This way of understanding complex interactions makes it possible to visualize and analyze complex spatio-ecological systems as a whole from individual interactions to local and global biogeographical properties.\\

\noindent\textbf{2. Delineating bioregions.} To identify bioregions we projected our biogeographical bipartite network on a spatial template (Step 2 in Figure \ref{Fig2}), by defining a metric to measure the similarity of species composition between grid cells. Several measures based on beta diversity have been proposed to quantify the degree of (dis)similarity between grid cells, typically taking into account the number of shared species between grid cells \citep{Wilson1984, Koleff2003}. These measures are mostly based on presence-absence data and aim at quantifying species turnover and species nestedness among grid cells, together or separately \citep{Baselga2012}. Although this indicator may be influenced by gradients in species richness \citep{Lennon2001,Baselga2012,Dapporto2015}, results obtained with the Jaccard index were more spatially coherent in our case. 

The resulting network is a weighted undirected spatial network whose intensity of links between grid cells range from 0, absence of a link (no species in common) to 1 (identical species composition). The detection of community structure in biogeographical networks is an interesting alternative approach to delineating bioregions \citep{Kougioumoutzis2014,Vilhena2015}. Community structure is indeed an important feature, revealing both the network internal organization and similarity patterns among its individual elements. In this study we used the Order Statistics Local Optimization Method (OSLOM) \citep{Lancichinetti2011}. OSLOM uses an iterative process to detect statistically significant communities with respect to a global null model (i.e. random graph without community structure). The main characteristic of OSLOM is that it is based on a score used to quantify the statistical significance of a cluster in the network \citep{Lancichinetti2010}. The score is defined as the probability of finding the cluster in a random null model. The random null model used in OSLOM is the configuration model \citep{Molloy1995} that generates random graphs while preserving an essential property of the network: the distribution of the number of neighbors of a node (i.e. the degree distribution). Therefore, the output of OSLOM consists in a collection of clusters that are unlikely to be found in an equivalent random network with the same degree sequence. This algorithm is nonparametric in the sense that it identifies the statistically significant partition, without defining the number of communities \textit{a priori}. However, the \textit{tolerance} value that determines whether a cluster is significant or not might play an important role for the determination of the clusters found by OSLOM. The influence of this value, fixed initially, is however relevant only when the community structure of the network is not pronounced. When communities are well defined, as it is usually the case in biogeography, the results of OSLOM do not depend on the particular choice of \textit{tolerance} value \citep{Lancichinetti2011}. See \citet{Lancichinetti2011} for a comparison between OSLOM and other community detection algorithms.

\noindent\textbf{3. Test-value matrix.} To analyse the bioregions and their species composition, we rely on \textit{test-values} measuring the under- or over-representation of species in a bioregion. Let us consider a studied area divided into $n$ grid cells, a species $i$ present in $n_i$ grid cells and a biogeographical region $j$ composed of $n_j$ grid cells. The test-value compares the actual number of grid cells $n_{ij}$, located in biogeographical region $j$ and supporting species $i$, with the average number $n_i n_j/n$ that would be expected if the species were uniformly distributed over the whole studied area. Since this quantity depends on $n_i$ and $n_j$ it is normalized by the standard deviation associated with the average expected number of grid cells \citep{Lebart2000}. The test value $\rho_{ij}$ is then defined as,

\begin{equation}
\displaystyle \rho_{ij}=\frac{n_{ij}-\frac{n_i n_j}{n}}{\sqrt{\frac{n-n_j}{n-1}\left(1-\frac{n_j}{n}\right)\frac{n_i n_j}{n}}} \label{rho}
\end{equation}

The test value $\rho_{ij}$ is negative if the species $i$ is under-represented in region $j$, equal to $0$ if the species $i$ is present in region $j$ in the same proportion as in the whole study area or positive if the species $i$ is over-represented in region $j$. In the latter case we consider that the species $i$ contribute positively to region $j$ and the level of contribution depends of the $\rho_{ij}$ value. Additionally, we consider that a plant species contribute positively and significantly to a bioregion $j$ if $\rho_{ij}$ is higher than a predetermined significance threshold $\delta$. Hence, The test-value matrix $\rho$ can be used to highlight set of species which better characterize the bioregions. The test-values are easy to interpret by specialists and represent an user-friendly way of ranking species according to their relevance.\\ 

\noindent\textbf{4. Groups of species.} The next step is to identify how similarities between species are spatially distributed across the study area. Here also we build a network in which the similarity $s_{ii'}$ between two species $i$ and $i'$ is equal to,

\begin{equation}
\displaystyle s_{ii'}=\frac{1}{1+\sqrt{\sum_j (\rho_{ij}-\rho_{i'j})^2}}
\end{equation}

This similarity metric is based on the Euclidean distance between test-values for each pair of species. Again, the community detection algorithm OSLOM is used to detect significant groups of species sharing the same spatial features (Step 4 in Figure \ref{Fig2}). This step produces a preliminary delimitation of the relationships between bioregions by identifying how the groups of species contributes to one or several bioregions.\\

\noindent\textbf{5. Coarse-grained biogeographical network.} To quantitatively characterize relationships between bioregions, we retained only the positive and significant species contributions by considering only test-values higher than $\delta=1.96$ (5\% significance level of a Gaussian distribution).

\begin{equation}
\displaystyle \rho_{ij}^{+}=\rho_{ij}\mathds{1}_{\rho_{ij}>1.96} \label{rhop}
\end{equation}

Then, since we are interested in interactions between bioregions we focused on the way species contributions are distributed among regions by normalizing $\rho^{+}$ by row (Equation \ref{chaprhop}).
\begin{equation}
\displaystyle \hat{\rho}_{ij}^{+}=\rho_{ij}^{+}/\sum_k \rho_{ik}^{+} \label{chaprhop}
\end{equation}

We then determined for each bioregions $j$ how the set of species $A_j=\left\{i\,|\,\rho_{ij}>1.96 \right\}$ that contributes to this biogeographical region are specific to it or also contribute to other regions (Equation \ref{lambda}).

\begin{equation}
\displaystyle \lambda_{jj'}=\frac{1}{|A_j|} \sum_{i \in A_j} \hat{\rho}_{ij'}^{+} \label{lambda}
\end{equation}

$\lambda_{jj'}$ represents therefore the average fraction of contribution to cluster $j'$ of species that contribute significantly to cluster $j$. The \emph{specificity} of a biogeographical region is therefore measured with $\lambda_{jj}$, while the \emph{relationships} with other regions is given by $\lambda_{jj'}$. It is important to note that for a given region $j$ the vector $\lambda_{j.}$ sum to one and can be expressed in percentage.\\

At the end of the process, we obtain a coarse-grained biogeographical network summarizing the biogeographical structure of the study area. This network is composed of the bioregions and the species groups (Step 5 in Figure \ref{Fig2}). All the metrics used to measure the similarity between the different bioregions are derived from the matrix of test-value $\rho$.

\begin{figure}[!h]
	\begin{center}
		\includegraphics[width=\linewidth]{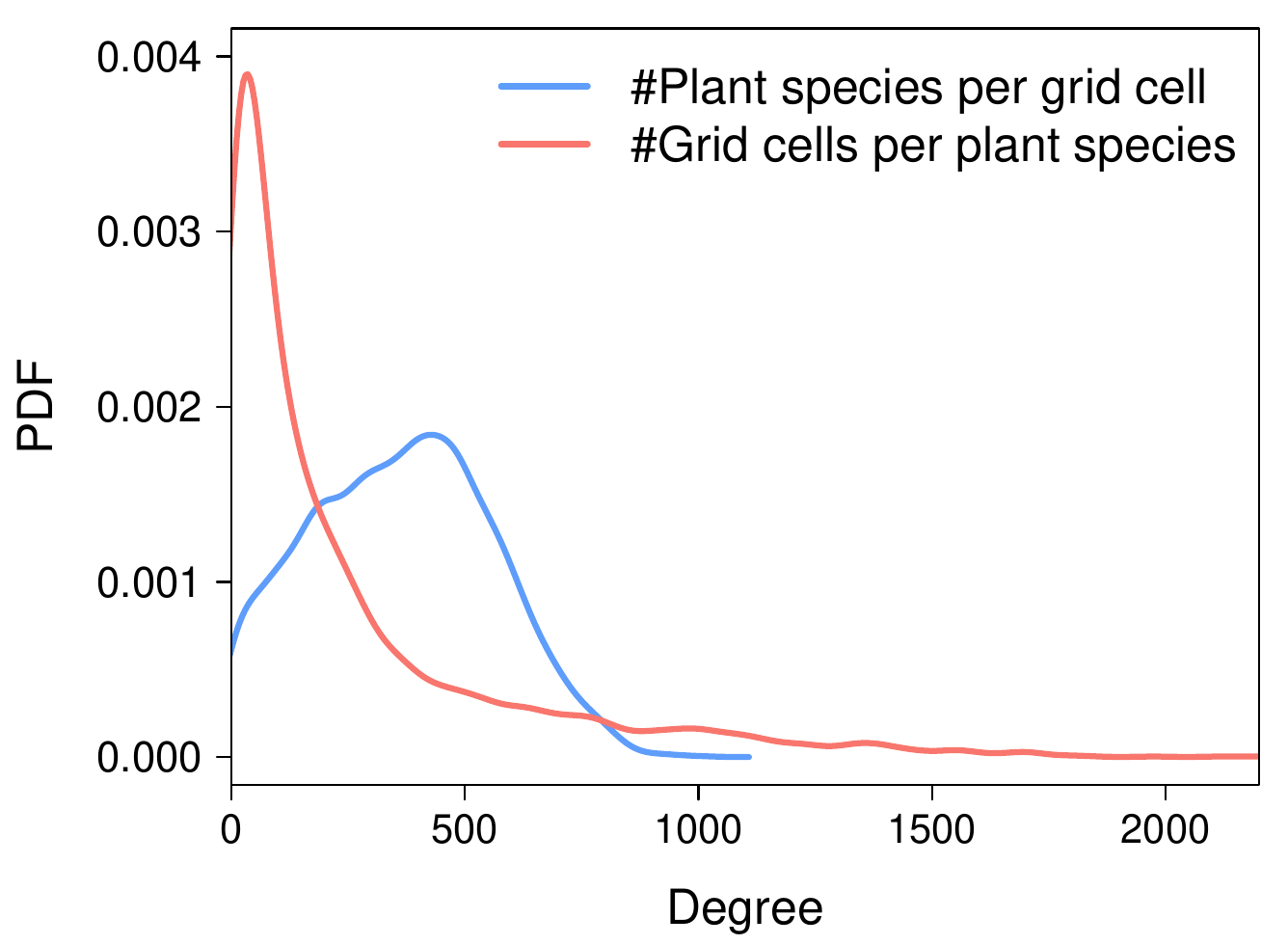}
		\caption{\sf \textbf{Degree distributions of the biogeographical bipartite network.} Probability density functions of the number of plant species per grid cell (in blue) and the number of cells covered per plant species (in red). Similar figures showing histograms instead of densities are available in Figure S\ref{FigS12} in Appendix. \label{Fig3}}
	\end{center}
\end{figure}  

\begin{figure*}
	\begin{center}
		\includegraphics[width=\linewidth]{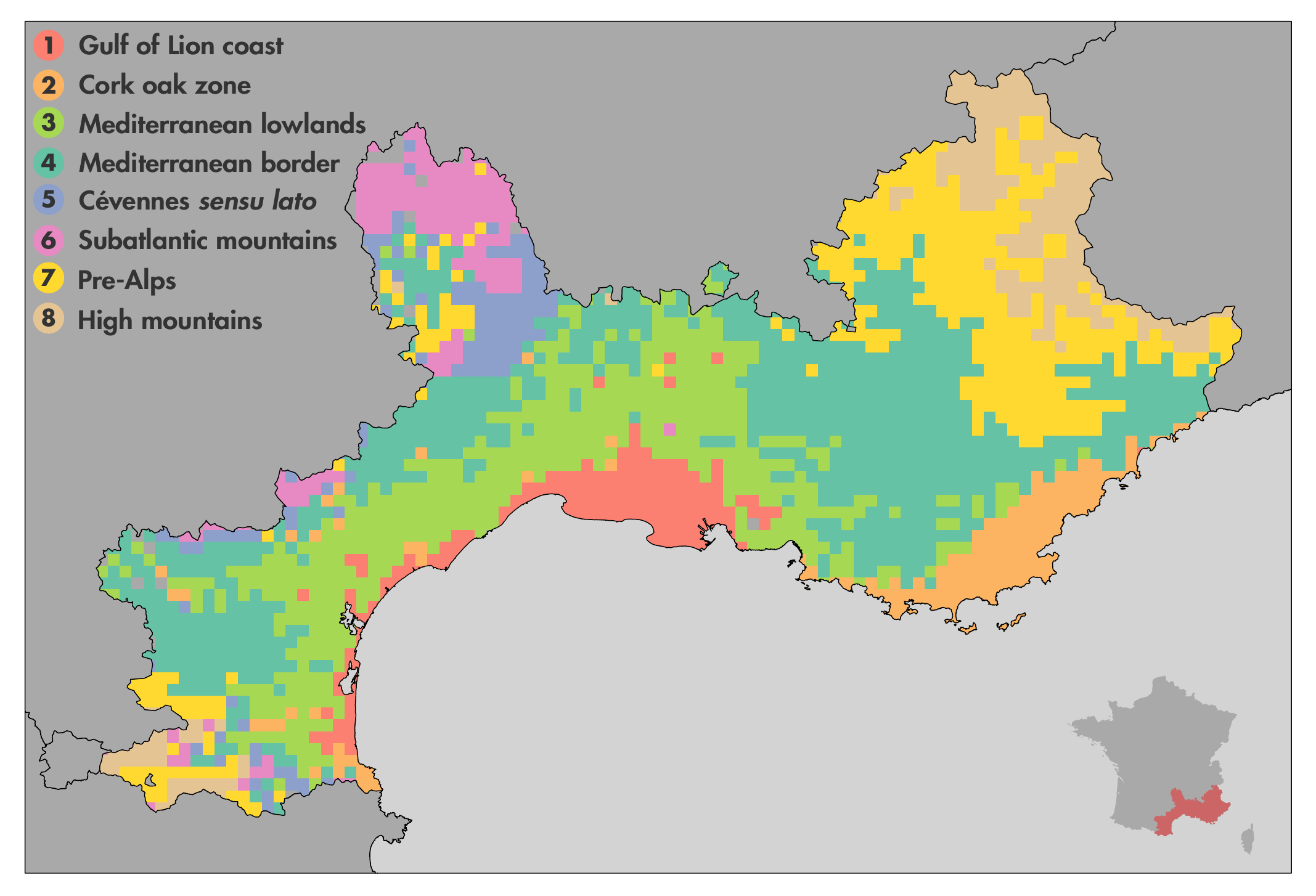}
		\caption{\sf \textbf{Bioregions based on similarity in plant species (l = 5 km).} Eight bioregions have been identified. 1. Gulf of Lion coast in red. 2. Cork oak zone in orange. 3. Mediterranean lowlands in light green. 4. Mediterranean border in dark green. 5. C{\'e}vennes \emph{sensu lato} in purple. 6. Subatlantic mountains in pink. 7. Prealps and other medium mountains in yellow. 8. High mountains in brown. The inset shows a map of France including the studied areas colored in red. An altitude map of the studied area is available in Appendix (Figure S\ref{FigS13}).  \label{Fig4}}
	\end{center}
\end{figure*}

\vspace*{-0.5cm}

\section*{Results}

\subsection*{Biogeographical bipartite network}

The bipartite network extracted from the database is composed of $2,607$ $5 \times 5$ km$^2$ grid cells and $3,697$ plant species, where the links represent the occurrence of plant species in the grid cells. Two network degree distributions can be associated to this network: the number of species per grid cell and the number of cells covered by each species. The probability density functions of these two distributions are displayed in Figure \ref{Fig3}. The spatial component of the network is very dense. Most of the grid cells host between 200 and 500 plant species, with an average of 360 species per cell (i.e. $\sim15$ species/km$^2$). For species side, the situation is different; the majority of plant species cover less than 10\% of the study area, which highlight the importance of range restricted taxa. Nevertheless, the distribution exhibits a long tail with a non-negligible number of widespread species. 

\begin{figure}
	\begin{center}
		\includegraphics[width=\linewidth]{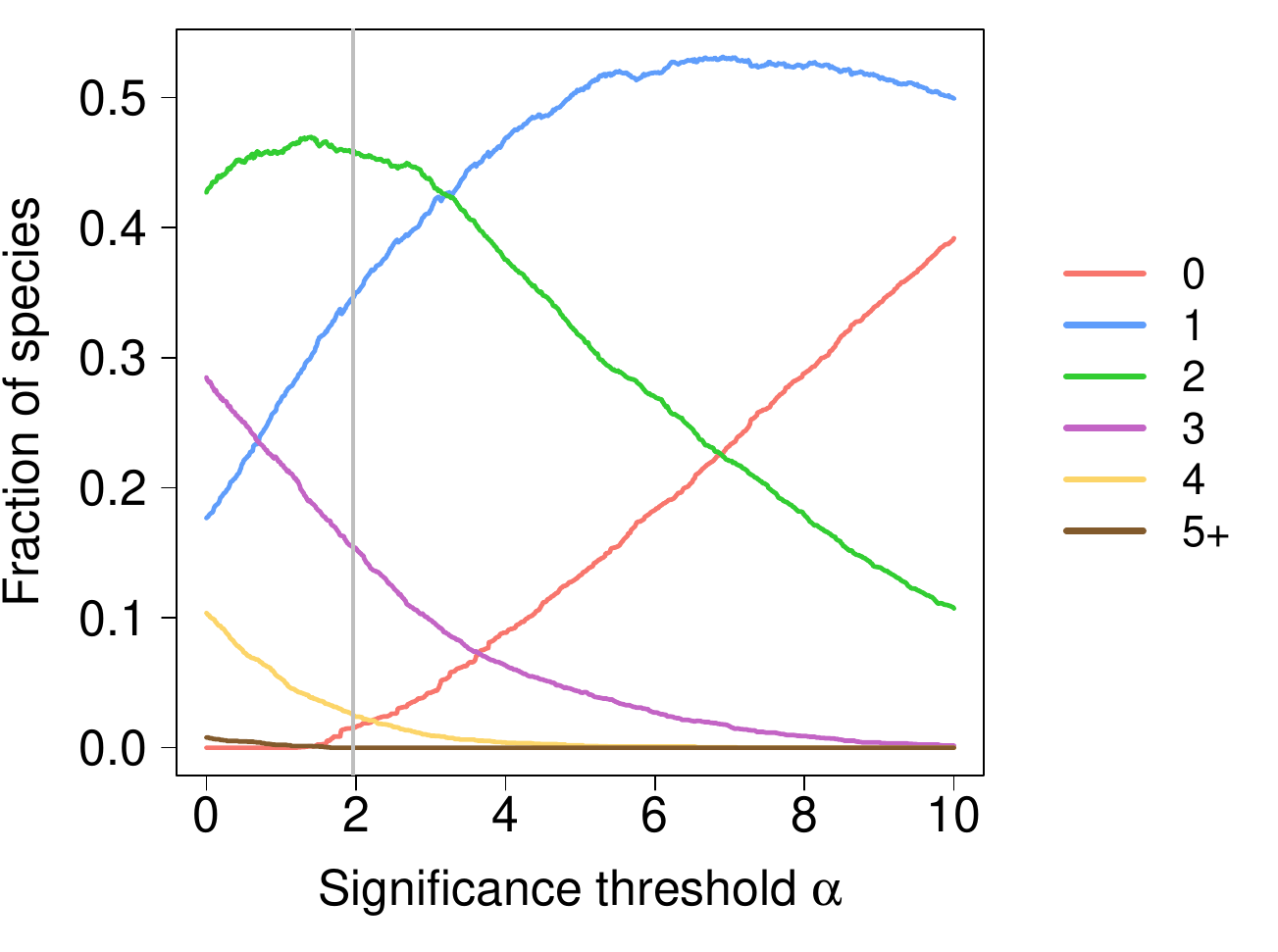}
		\caption{\sf \textbf{Fraction of species contributing positively and significantly to a given number of bioregions (from 0 to 5 or more) as a function of the significance threshold.} The vertical line represents the significance threshold $\delta=1.96$. \label{Fig5}}
	\end{center} 
\end{figure}

\subsection*{Delineating bioregions}

We identified eight statistically significant bioregions reflecting the biogeographical structure of the French Mediterranean area based on plant species distribution (Figure \ref{Fig4}). Clusters size vary from 120 to 807 square cells. Clusters are spatially coherent, exhibiting a connectivity measure always higher than $0.5$ (i.e. ratio between the number of grid cells in the largest patch and the total number of grid cells \citep{Turner2001}). Results obtained are not scale sensitive, and the spatial coherence of each cluster according to the scale ($l=5$ and $10$ km) can be found in Table S\ref{TabS1} in Appendix. It also important to note that this step can also be performed with standard hierarchical clustering methods. The results obtained with Ward's clustering are available in Appendix.

\subsection*{Groups of plant species}

The test-value matrix can be used to identify plant species that contribute positively and significantly to one or more bioregions. It is worth noting that the number of contributions and their intensity vary among species. Indeed, some species contribute very little to only one region while other species contribute significantly to three or more regions. The number of species contributing to a given number of regions depends on the significance threshold $\delta$. A very small and negative value of $\delta$ will imply that almost all plant species contribute significantly to the 8 bioregions. In contrast, a very high value of $\delta$ will result in all species contributing to no regions. In order to get a better understanding of species contribution mechanisms and to assess the influence of $\delta$, we plot in Figure \ref{Fig5} the fraction of species contributing positively to a given number of bioregions as a function of a significance threshold value. If we consider the default threshold $\delta=1.96$, that corresponds to a 2.5\% significance level of a Gaussian distribution, we observe that the vast majority of plant species contributes positively to one or two regions representing 35\% and 45\% of species, respectively. There is also 20\% percent of plant species that contribute to three or more bioregions. If we increase the minimum level of contribution necessary to claim that a species contributes to a region, we see that the fraction of species contributing to two or more bioregions dramatically decreases while the fraction of species with no contribution increases. However, it is interesting to note that the fraction of species contributing to one region to increases until reaching a plateau. This demonstrates that 50\% of plant species are strongly connected to a single region.  

The similarities between plant species' contribution to the 8 regions allowed us to identify 20 groups of species, and their contribution to each bioregion is displayed in Figure \ref{Fig6}. We observed different patterns of contributions in terms of shape and intensity. This allows for the identification of groups of species sharing similar spatial features and highlights relationships between bioregions through the way plant species contribute to different group of regions.     

\subsection*{Relationships between bioregions}

\begin{figure}[!h]
	\begin{center}
		\includegraphics[width=\linewidth]{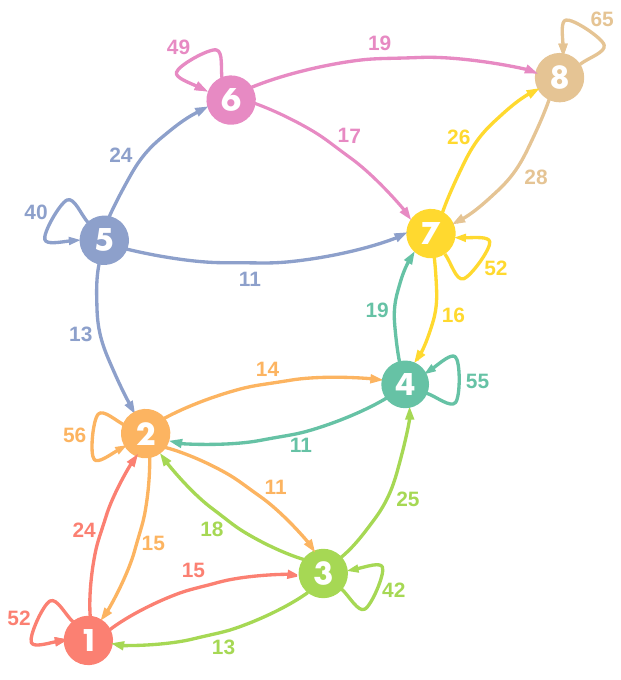}
		\caption{\sf \textbf{Network of interactions between bioregions.} $\lambda_{jj'}$, expressed here in percentage, represents the average fraction of contribution to cluster $j'$ of species that contribute significantly to cluster $j$. Only links with a value $\lambda_{jj'}$ higher than 10\% are shown.\label{Fig7}}
	\end{center}
\end{figure}

\begin{figure*}
	\begin{center}
		\includegraphics[width=\linewidth]{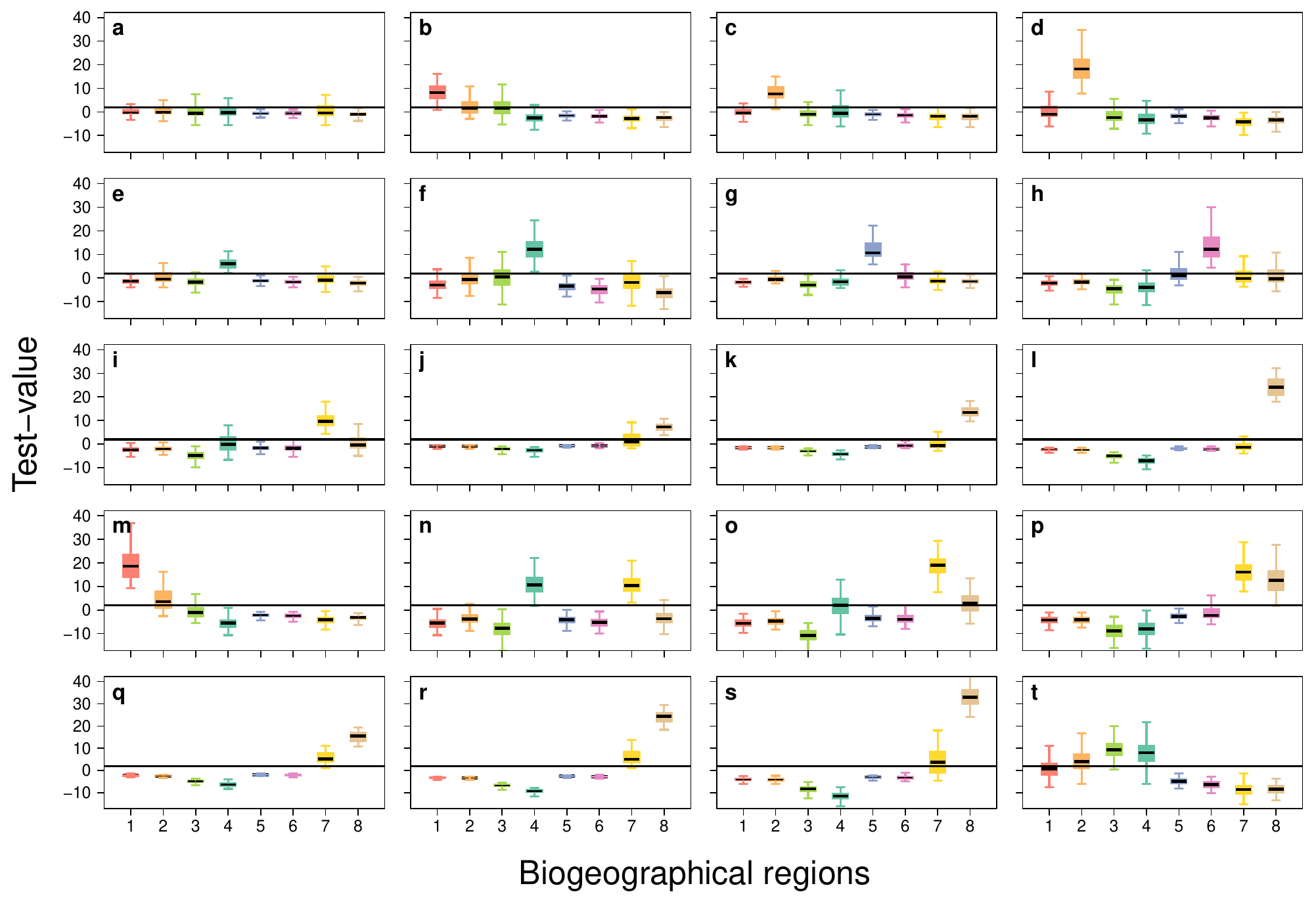}
		\caption{\sf \textbf{Description of the groups of plant species.} Boxplot of test-values according to the bioregions and the plant species groups. The horizontal line represents the significance threshold $\delta=1.96$. The number of plant species per group is available in Table S\ref{TabS3} in Appendix. \label{Fig6}}
	\end{center} 
\end{figure*}

This leads us to the study of relationships between bioregions. The network of interactions $\lambda$ derived from the test-value matrix is plotted in Figure \ref{Fig7}. We found that, globally, plant species contributing significantly to a region contribute mostly to this region, with an average specificity of 51\% across the eight bioregions. It must be pointed out however that some regions are more specific than others with $\lambda_{jj}$ values ranging from 40\% to 65\%. 

Analysis of how bioregions connect with each other showed that there is no isolated region in the sense that every region is connected with at least one other region with a $\lambda_{jj'}$ value varying from 1 to 28\%. Moreover, for all regions, the maximal $\lambda_{jj'}$ value is always higher than 10\%. Although it is generally the case, it is also worth mentioning that the relationships are not necessarily symmetric, which represents an interesting way of detecting hierarchical relationships. A table displaying all $\lambda_{jj'}$ values is available in Table S\ref{TabS4} in Appendix.

\vspace*{-0.5cm}

\section*{Discussion}

In this study we delineate spatial bioregions in southern France, a transition area between a mediterranean and temperate climate. The present analysis represents to our knowledge one of the largest network-based studies published to date, relying on a database containing more than four million data points across a territory of about $558,776 \mbox{ km}^2$. While this territory has been divided into bioregions on expert knowledge, we confront those approaches to data-driven classification, and discuss the coherence of the different perspectives. We delineated eight statistically significant bioregions, which we will first present in relation to previously published work, and emphasize their specificity regarding associated groups of species. We discuss the observed spatial patterns in terms of ecological and historical drivers, to provide insights into mechanisms driving the assemblage of vegetation communities. 

\subsection*{Bioregions}

The clustering approach identified eight statistically significant spatial clusters, that represents coherent territories detailed below. Regions are presented from Mediterranean toward temperate and mountainous climates.\\

\noindent \textbf{1. Gulf of Lion coast} is a bioregion that extends west of the Rh{\^o}ne, penetrating more inland around the wetland of the Rh{\^o}ne Delta. The latter, along with the Languedoc lagoons, is frequently used as an example of azonal vegetation \citep{Ozenda1994}, and the originality of the flora and the vegetation of these areas has long been recognized \citep{Molinier1970}. Some subdivisions have been suggested separating, even at a coarse scale, the sand-dune complex, the halophytic vegetation and the salt meadows \citep{Bohn2000}, but were not found here probably due to the size of the cells we used. From a geological point of view, this bioregion is essentially made of sand dunes, lagoon sediments and modern alluvium. It is entirely situated under a Mediterranean climate, in the mesomediterranean climatic belt, with a dry season of two or three months in the summer \citep{Rivas2004a}. Taxa specific to this cluster exhibit a distribution following the Mediterranean coastal area, extending in some cases towards other coastal areas or to arid inland zones. They are mostly encountered in halophytic communities and surprisingly not that much into dunes, suggesting that the key factor defining this bioregion might be the saline soils rather than the coastal position alone. Several narrow-endemics rely on those habitats, especially in the genus Limonium whose rapid radiation is typical of mediterranean neoendemics \citep{Lledo2005}.\\

\noindent \textbf{2. Cork oak zone} encompasses the Maures-Est{\'e}rel range and neighbouring areas. West of the Rh{\^o}ne, it is fragmented with cells in the eastern tip of the Pyrenees (low Alb{\`e}res and the Roussillon lowlands), plus a few more sparsely dispersed zones in Languedoc. The Provence and Alb{\`e}res areas have been identified by phytogeographers \citep{Ozenda1987, Ozenda1994} as the \enquote{Cork oak zone}, a silicicolous warm mesomediterranean area. Indeed, climatic data show a clear summer dry period of one to two months. Almost all of the cells contain acidic soils over a variety of substrates (granites, gneiss, schists, sandstones, alluvial deposits, etc.). Species most linked to the \enquote{Cork oak zone} have a Mediterranean distribution, with some extending towards the Atlantic area. Characteristic species have ecological preferences for acid soils, and belong to various vegetation stages (forest, scrub or grassland formations).\\

\noindent \textbf{3. Mediterranean lowlands} bioregion covers the hinterland of the Gulf of Lion from the Roussillon to western Provence. Several authors have individualized an arc shaped mesomediterrean zone \citep{Dupias1985,Ozenda1994} but their limits do not fit exactly ours. The closest match is the catalonian-proven{\c c}al mesomediterranean holm oak forests unit of the European natural vegetation map \citep{Bohn2000}. The area is principally composed of sedimentary rocks (mostly limestones and marls) and alluvium. Its climate is Mediterranean \citep{Rivas2004a}, with a summer dry period of one to three months. With few exceptions, species most linked to this bioregion have a distribution included in the Mediterranean region \citep{Rivas2004}. Most of them belongs to communities of the \emph{Quercetea ilicis} or of the former \emph{Thero-Brachypodietea}, i.e. the matorral / forest and grasslands communities making up the landscape locally called \enquote{garrigues}. The other part of these taxa is usually found in disturbed communities, showing the strong incidence of human activities in this area.\\

\noindent \textbf{4. Mediterranean border} is a bioregion whose northern edge roughly follows the limit of the Mediterranean world as it is usually depicted \citep{Dupias1985,Quezel2004}. It broadly coincide to what has been called a supramediterranean belt \citep{Ozenda1994} or a submediterranean zone \citep{Bolos1961}, and fits quite well with four mapping units of the Map of the natural vegetation of Europe \citep{Bohn2000}; namely the catalonian-proven{\c c}al supramediterranean holm oak forests and three types of downy oak forests (ligurian- middle apennine, languedocian and those extending from the southern Pyrenees to the southwest pre-Alps). The substratum of this area is mainly calcareous and marly. This area  has a short (one month) summer drought period with the exception of some Var and Alpes-Maritimes places where the summer drought is more pronounced (two months). Species most linked to this bioregion present a western eury-mediterranean distribution, and share a common ecology, occurring frequently in communities belonging to the \emph{Helianthemo italici-Aphyllanthion monspeliensis} and to a lesser extent to the \emph{Ononidetalia striatae} \citep{Gaultier1989,Rivas2002}, i.e. dry dwarf scrubs and their associated grasslands on calcareous and marly eroded soils \citep{Mucina2016}.\\

\noindent \textbf{5. C{\'e}vennes \emph{sensu lato}} is a bioregion to which most of the cells are situated in the C{\'e}vennes areas, while the remainder is scattered over the eastern Pyrenees piedmont and the Montagne Noire (southern limit of the Massif Central). This spatial cluster overlays four zones of the phyto-ecological regions \citep{Dupias1985}: the lower C{\'e}vennes, the \enquote{warm} C{\'e}vennes valleys, the Aspres and the chestnut zone of the southern edge of the Montagne Noire. The C{\'e}vennes proper part of this cluster has also been identified by other authors \citep{Braun1923,Ozenda1994} and putative glacial refugia has been positioned there \citep{Medail2009}. This area is not subject to a summer drought and covers siliceous substrata such as schists, granites or gneiss. Taxa exhibiting the strongest link to this biogeographical region are either C{\'e}vennes endemics, subendemics \citep{Lavergne2004, Dupont2015} or plants with a more or less Atlantic distribution \citep{Dupont2015}, but no clear ecological pattern is emerging among these taxa.\\

\noindent \textbf{6. Subatlantic mountains} The largest area covered by cells of this biogeographical region is the northern part of the Loz{\`e}re department. The remaining cells are mostly distributed in the Massif Central and in the Pyrenees. These areas belong to the beech (Fagus sylvatica L.) montane belt \citep{Ozenda1994, Bohn2000} with a few exceptions where Scots pines (Pinus sylvestris L.) dominate. It corresponds to the predominantly siliceous subatlantic type \citep{Ozenda1987}, where the climate is rather wet, with precipitations frequently exceeding 1,000 mm per year and no dry period. Thus, wetlands and bogs are not rare, and the substratum is made of igneous rocks which explain the acidic nature of the soils. The majority of the taxa most linked to this spatial cluster are generally distributed all over the eurosiberian region or the western part of this region, corresponding to a subatlantic distribution \citep{Rivas2004, Dupont2015}. Interestingly, most of those plants grow in wetlands habitats, a trend already noticed in the Massif Central \citep{Braun1923}.\\

\noindent \textbf{7. Pre-Alps and other medium mountains} represent a bioregion whose cells are disseminated through the lower parts of the eastern Pyrenees including almost all the Pyrenean part of the Aude department, through the highest areas of the Causses, around the Mont Ventoux and through the most eastern part of the Pre-Alps. This area has rarely been individualised in such a way even if at a European scale it can be related to several more or less calcicolous beech or fir-beech forest belts \citep{Bohn2000} (Abies alba L. and Fagus sylvatica L.), or more specifically, for the Var department, to a pre-alpine district \citep{Lavagne2008}. Most of the rock underlying this area is calcareous. Climatically, we are outside of the Mediterranean climate as there is no dry period. The distribution of taxa most linked to this biogeographical region is basically holarctic, avoiding the Mediterranean parts of Europe. Some of these taxa also avoid the most Atlantic part of the continent. Their ecology is varied, pertaining to different stages (grasslands, shrubs, forests) of mountain vegetation series, often (but not systematically) calcicolous.\\

\noindent \textbf{8. High mountains} This bioregion regroups the highest part of the Alps and the Pyrenees. If most authors agree on individualizing the upper vegetation belts of these mountain ranges, its unity and the common points are less often identified \citep{Ozenda2002}. Both calcareous and acidic soils are to be found in this area. Cells of this region are the coldest of our study area, and there is no dry period: the climate is relatively harsh and the vegetation period is reduced \citep{Ozenda2002} compared to the other clusters. Taxa most linked to this region are mainly European mountains endemics, venturing also in the Arctic. They belong to grasslands or snowbeds communities, which is consistent with their occurrence on the highest ranges.\\

\subsection*{Species and spatial relationships among bioregions}

\subsubsection*{Defining the Mediterranean region} 

At a global scale, the delimitation of the Mediterranean border has been a long running question \citep{Latini2017}, and the mismatch of the numerous attempts attest to the difficulties (Figures S\ref{FigS4}-\ref{FigS11} in Appendix). In France, the first attempt goes back to third edition of the Flore Fran{\c c}aise by \citet{Lamarck1805}, as shown in \citet{Ebach2006} followed by several other works such as \citet{Flahault1887}, who considered the distribution limit of the olive tree (Olea europaea L.) as a marker of the Mediterranean biome. This was later generalized to the evergreen oak belt \citep{Quezel1999}, but it appeared that the situation was more complex \citep{Quezel2004}. Thus, variability in results has not lead to a comprehensive framework yet. This has several implications regarding conservation programs, as the delimitation mentioned by European legislation has been used as a reference to delimit the distribution of several protected habitat\footnote[2]{\url{http://www.eea.europa.eu/data-and-maps/F50D9CF8-FFEE-475F-8A65-4E095512CBB7} (accessed the 04/07/2018)}. In this study, the network approach allowed to discriminate two \enquote{sub-networks} with little exchange regarding species composition and different relative contribution to each area, which globally relate to a temperate and a Mediterranean sub-groups. Several earlier bioregionalizations in the mediterranean basin have failed to separate mediterranean from eurosiberian ensembles, suggesting this boundary would be highly permeable  \citep{Saiz1998,Garcia2002} and easily crossed by species. Here, the use of a precise dataset coupled with a network analyse has proven to be relevant to depict such spatial transition, which reinforce the need to gather coherent dataset to characterize complex and intricate spatial structures. This biogeographical boundary has been linked to a change in the annual distribution of precipitation, which induces a prolonged summer drought and a stronger climatic seasonality in the mediterranean \citep{Antonelli2017}. At a finer scale, the three mediterranean clusters present a high spatial coherence, and closely fit to the mesomediterranean thermoclimatic belt \citep{Rivas2004} (see Figure S\ref{FigS10} in Appendix). The high congruence between climatic model \citep{Rivas2004} and biogeographic entities has never been pictured by previous bioregionalization works (see Appendix for maps), as most of them presented a wider definition of the mediterranean biome, extending northward. 
Then, the absence of orogenic barriers along this climate-based distinction is likely to produce shallow boundaries typical of transition areas \citep{Antonelli2017, Ficetola2017} exemplified here by the cluster \enquote{Mediterranean border} that contains all historical attempt to delimitate the mediterranean biome. West of the Rhone, this region is relatively thin and fence around the mesomediterranean ensemble; east of the Rhone, it occupies a wide area on the Alpine piedmont.  Thus, instead of drawing a single line \citep{Cox2001}, we propose to identify a transition area \citep{Latini2017,Droissart2017} with an upper boundary as the limit of the Mediterranean biome \citep{Antonelli2017}.

\subsubsection*{Vicariance and fragmentation among bioregions} 

The relationship between bioregions can be seen through the understanding of species relative importance in each area. First, the regions \enquote{Gulf of Lion coast}, \enquote{Cork oak zone} and \enquote{Mediterranean lowlands}, all included within the same bioclimatic belt \citep{Rivas2004a}, differ mostly on substratum, i.e. calcareous (bioregion 3), siliceous (bioregion 2) or quaternary deposits (bioregion 1). Thus, they are well defined and little uncertainty exists concerning their spatial configuration (Figure S\ref{FigS14} in Appendix); those three entities can be seen as climatic vicariant bioregions which have conjointly developed on different geological substrates, or \enquote{islands}. As a result, they share an important pool of species, and present the highest complementarity in the network, as they are the only three clusters all related to each other. In contrast, the relationship between the \enquote{Cork oak zone} and the \enquote{C{\'e}vennes} exemplify the opposite process: those two areas share a similar bedrock (mainly acidic substrate) but are located at each extreme of the Mediterranean climatic gradient. While the \enquote{Cork oak zone} is present under hot and dry mesomediterranean climate (some coastal cells even belonging to a thermomediterranean belt), the \enquote{C{\'e}vennes} present a higher impluvium and a very weak summer drought. Consequently, they share a common set of species which interestingly are typical of the \enquote{C{\'e}vennes} cluster, and extend into the \enquote{Cork oak zone}. Noteworthy point, those population can constitute relictual rear edge populations, which often retain particular interest for conservation \citep{Hampe2005, Lavergne2006}. 

Finally, the \enquote{Pre-Alps} and \enquote{High mountains} bioregions are both present within the three mountain chains, and occupy climatic conditions with no dry period at all, and especially harsh prolonged winter for the second. Several species groups are highly informative for both of those bioregions, which signify that they share an important group of species globally adapted to mountain environment. \enquote{High mountains} present the highest percentage of typical species. Yet, within the numerous plant species groups characterizing those entities (5 groups in Figure \ref{Fig6}), the relative contribution of each toward one or the other bioregion might differ slightly, sometimes in association with another bioregion such as the \enquote{Mediterranean border} (Figure \ref{Fig6}). This illustrates that groups of taxa are unevenly important across these two regions, probably reflecting the complex geological substrate. Thus, while our analysis reflect an overall homogeneity of mountain flora mainly driven by climate, it is likely that finer divisions based on a more precise study could be expected. This has been pinpointed by \citet{Bohn2000} who pictured a high local heterogeneity due to steep altitudinal gradients and geological diversity, despite some vegetation groups shared between the Alps and the Pyrenees. Therefore, a comparative analysis including a broader spatial perspective on those massif could improve our understanding of the spatial structure of mountain flora in western Europe.

\subsubsection*{Eco-evolutionary factors driving the spatial organisation of plant diversity} 

The spatial distribution and species relative importance for each bioregion can help us to better understand processes that have shaped Mediterranean biota in the south of France. The regional species pool results from several waves of colonization following glacial cycles, constrained by ecological filters that allowed taxa to persist and ultimately shaped local communities \citep{Ricklefs1987}. Indeed, our study area is at the crossroad of recolonization routes out of two major refugia, i.e. the Iberic and Italian peninsulas \citep{Hewitt2000}, and represents an admixture zone for several mediterranean taxa \citep{Lumaret2002}. Joint action of colonization-retraction sequences and long term persistence within microrefugia has been suspected to generate fragmented distribution. Thus, one particular feature of such climatic transition area is the high proportion of population isolated at the periphery of their main range \citep{Thompson2005a}, either at the rear or at the leading edge of their distribution \citep{Hampe2005}. However, spatial patterns alone do not inform on the evolutionary isolation of such populations, could it be of recent dispersal following Last Glacial Maximum \citep{Lumaret2002}, or long term persistence in a given refugia \citep{Medail2009, Papuga2015}. Thus, integrating phylogenies within bioregionalization would prove informative to analyse historical events that have shaped current spatial patterns of biodiversity \citep{Nieto2014}, and capture the evolutionary relationship among bioregions \citep{Holt2013}.

Nevertheless, analysing the spatial organization of flora can help us to understand ecological factors that shape such bioregions. Orographic barriers and past tectonic movement are expected to have little impact on our study area, as no such events have occurred there since the onset of the Mediterranean climate in the Pliocene \citep{Rosenbaum2002}. In our analysis, spatial structuration relies principally on two elements. On the one hand, a climatic gradient from Mediterranean to temperate climate creates fuzzy spatial limits among adjacent groups, and increases uncertainty when delimitating groups (Figure S\ref{FigS14} in Appendix). This is exemplified by the spatial imbrication of \enquote{Mediterranean lowlands} and \enquote{Mediterranean border}. On the other hand, geological variations can form sharp transitions creating important species turnover between places close apart. This is exemplified by the \enquote{Cork oak zone} whose spatial delimitation is very clear, due to the presence of an acidic substrate surrounded by places dominated by calcareous-based rock. Interestingly, this area still shares an important part of its biota with other places in the Mediterranean basin probably inherited from times where such geological islands formed a single ensemble, before the separation and later migration of these islands \citep{Rosenbaum2002, Medail1997}. The joint action of these two ecological factors has already been highlighted in previous bioregionalization of the Mediterranean basin \citep{Buira2017}. As a result, complex geo-climatic variation have played a key role in shaping island-like territories which have fragmented species distributions, a factor that has strong influence on populations characteristics both genetically and demographically \citep{Pironon2017}.

The flora of the Mediterranean basin shows recurrent patterns of narrow endemism, species turnover and  highly disjunct distributions \citep{Thompson2005a}. While allopatric isolation has been suspected to be the main mechanism explaining the differentiation of taxa, the shared significance of different ecological variables (namely climate and geology) points out the combined importance of spatial isolation and heterogeneous selective pressures \citep{Thompson2005a,Anacker2014}. Additionally, recent studies have shown that this can be enhanced by small scale changes of the ecological niche \citep{Papuga2018,Lavergne2004, Thompson2005b}. Contrary to other mediterranean biomes (e.g. South-Africa and Australia), the mediterranean basin is marked by an active speciation, which has led to the high observed proportion of neoendemic species \citep{Rundel2016}. If evidences have accumulated concerning cryptic microrefugia for temperate trees \citep{Stewart2001}, little is known regarding mediterranean taxa, especially those that exhibit little dispersal capacities, a shared trait among mediterranean endemics \citep{Lavergne2004}. Thus, this bioregionalization set the scene to investigate the shared phylogeographic legacy of the Mediterranean biota \citep{Puscas2012}, and measure the evolutionary isolation of such communities that separate peripheral isolates from newly differentiated species \citep{Crawford2010}.

\section*{Conclusion}

The quality of a bioregionalization is dependent on the data and the method used. To our knowledge, the present analysis constitutes the densest species-cells network analysed in a bioregionalization study, at such a high spatial resolution. Therefore, results of this study demonstrate that new statistical methods based on network analysis can bring solutions to manage and analyse large databases, and provide efficient bioregionalization at different scales. New perspectives for bioregionalization will integrate community structure across different scales, in order to understand how deterministic (i.e. niche based) processes and stochastic events (dispersal, random extinction, ecological drift) interact to shape plant communities, from regional species pool to local assemblages \citep{Chase2011}.

\section*{Acknowledgements}

This work was supported by a grant from the French National Research Agency (project NetCost, ANR-17-CE03-0003 grant). Partial financial support has been received from the French Minist{\`e}re de la Transition Ecologique et Solidaire (MTES). We thank the Alpine National Botanic Conservatory for providing some of the Provence-Alpes-C{\^o}te d'Azur data. We wish to thank Christelle H{\'e}ly-Alleaume, Virgile Noble and James Molina for useful discussions. A special thank goes to John D. Thompson for correcting English and interesting remarks.

\section*{Data availability}

Code and data are available at \url{www.maximelenormand.com/Codes}

\bibliographystyle{myapalike}
\bibliography{Biogeo}

\begin{thebibliography}{}

\bibitem[Anacker \& Strauss, 2014]{Anacker2014}
Anacker, B.~L. \& Strauss, S.~Y. (2014)\removeperiod.
\newblock The geography and ecology of plant speciation: range overlap and
  niche divergence in sister species.
\newblock {\em Proceedings of the Royal Society of London B: Biological
  Sciences}, \textbf{281}, 20132980.

\bibitem[Antonelli, 2017]{Antonelli2017}
Antonelli, A. (2017)\removeperiod.
\newblock Biogeography: {Drivers} of bioregionalization.
\newblock {\em Nature Ecology \& Evolution}, \textbf{1}, 0114.

\bibitem[Baselga, 2012]{Baselga2012}
Baselga, A. (2012)\removeperiod.
\newblock The relationship between species replacement, dissimilarity derived
  from nestedness, and nestedness.
\newblock {\em Global Ecology and Biogeography}, \textbf{21}, 1223--1232.

\bibitem[Blondel \textit{et~al.}, 2010]{Blondel2010}
Blondel, J., Aronson, J., Bodiou, J.-Y. \& Boeuf, G. (2010)\removeperiod.
\newblock {\em The {{Mediterranean Region}}: {{Biological Diversity}} in
  {{Space}} and {{Time}}}.
\newblock {Oxford University Press}, Oxford, New York, second edition edition.

\bibitem[Bohn \textit{et~al.}, 2000]{Bohn2000}
Bohn, U., Gollub, G. \& Hettwer, C. (2000)\removeperiod.
\newblock {\em Karte der nat{\"u}rlichen Vegetation Europas. Bundesamt f{\"u}r
  Naturschutz}.
\newblock Landwirtschaftsvlg M{\"u}nster, Bonn, 1., aufl. edition.

\bibitem[B{\'o}los, 1961]{Bolos1961}
B{\'o}los, O. (1961)\removeperiod.
\newblock La transici{\'o}n entre la depresi{\'o}n del ebro y los pirineos en
  el aspecto geobot{\'a}nico.
\newblock {\em Anal. lnst. Bot. Cavanilles}, \textbf{18}, 99--254.

\bibitem[Braun-Blanquet, 1923]{Braun1923}
Braun-Blanquet, J. (1923)\removeperiod.
\newblock {\em L'origine et le d{\'e}veloppement des flores dans le massif
  central de {France}; avec aperçu sur les migrations des flores dans
  l'{Europe} sud-occidentale}.
\newblock L. Lhomme, Paris.

\bibitem[Buira \textit{et~al.}, 2017]{Buira2017}
Buira, A., Aedo, C. \& Medina, L. (2017)\removeperiod.
\newblock Spatial patterns of the {Iberian} and {Balearic} endemic vascular
  flora.
\newblock {\em Biodiversity and Conservation}, \textbf{26}, 479--508.

\bibitem[Chase \& Myers, 2011]{Chase2011}
Chase, J.~M. \& Myers, J.~A. (2011)\removeperiod.
\newblock Disentangling the importance of ecological niches from stochastic
  processes across scales.
\newblock {\em Philosophical Transactions of the Royal Society of London.
  Series B, Biological Sciences}, \textbf{366}, 2351--2363.

\bibitem[Cheruvelil \textit{et~al.}, 2017]{Cheruvelil2017}
Cheruvelil, K.~S., Yuan, S., Webster, K.~E., Tan, P.-N., Lapierre, J.-F.,
  Collins, S.~M., Fergus, C.~E., Scott, C.~E., Henry, E.~N., Soranno, P.~A.,
  Filstrup, C.~T. \& Wagner, T. (2017)\removeperiod.
\newblock Creating multithemed ecological regions for macroscale ecology:
  {Testing} a flexible, repeatable, and accessible clusteringÂ method.
\newblock {\em Ecology and Evolution}, \textbf{7}, 3046--3058.

\bibitem[Cox, 2001]{Cox2001}
Cox, B. (2001)\removeperiod.
\newblock The biogeographic regions reconsidered.
\newblock {\em Journal of Biogeography}, \textbf{28}, 511--523.

\bibitem[Crawford, 2010]{Crawford2010}
Crawford, D.~J. (2010)\removeperiod.
\newblock Progenitor-derivative species pairs and plant speciation.
\newblock {\em Taxon}, \textbf{59}, 1413--1423.

\bibitem[Dahlin \textit{et~al.}, 2014]{Dahlin2014}
Dahlin, K.~M., Asner, G.~P. \& Field, C.~B. (2014)\removeperiod.
\newblock Linking vegetation patterns to environmental gradients and human
  impacts in a mediterranean-type island ecosystem.
\newblock {\em Landscape Ecology}, \textbf{29}, 1571--1585.

\bibitem[Dapporto \textit{et~al.}, 2015]{Dapporto2015}
Dapporto, L., Ciolli, G., Dennis, R. L.~H., Fox, R. \& Shreeve, T.~G.
  (2015)\removeperiod.
\newblock A new procedure for extrapolating turnover regionalization at
  mid-small spatial scales, tested on {{British}} butterflies.
\newblock {\em Methods in Ecology and Evolution}, \textbf{6}, 1287--1297.

\bibitem[Div{\'\i}{\v s}ek \textit{et~al.}, 2016]{Divisek2016}
Div{\'\i}{\v s}ek, J., Storch, D., Zelen{\'y}, D. \& Culek, M.
  (2016)\removeperiod.
\newblock Towards the spatial coherence of biogeographical regionalizations at
  subcontinental and landscape scales.
\newblock {\em Journal of Biogeography}, \textbf{43}, 2489--2501.

\bibitem[Droissart \textit{et~al.}, 2017]{Droissart2017}
Droissart, V., Dauby, G., Hardy, O.~J., Deblauwe, V., Harris, D.~J., Janssens,
  S., Mackinder, B., Blach-Overgaard, A., Sonk{\'e}, B., Sosef, M.~M.,
  St{\'e}vart, T., Svenning, J.-C., Wieringa, J.~J. \& Couvreur, T. L.~P.
  (2017)\removeperiod.
\newblock Beyond trees: {Biogeographical} regionalization of tropical {Africa}.
\newblock {\em Journal of Biogeography}.

\bibitem[Dupias \& Rey, 1985]{Dupias1985}
Dupias, G. \& Rey, P. (1985)\removeperiod.
\newblock {\em Document pour un zonage des r{\'e}gions phyto-{\'e}cologiques}.
\newblock Centre d'{\'e}cologie des ressources renouvelables.

\bibitem[Dupont, 2015]{Dupont2015}
Dupont, P. (2015)\removeperiod.
\newblock {\em Les plantes vasculaires atlantiques, les
  pyr{\'e}n{\'e}o-cantabriques et les {\'e}l{\'e}ments floristiques voisins
  dans la p{\'e}ninsule ib{\'e}rique et en {France}}.
\newblock Soci{\'e}t{\'e} botanique du Centre-Ouest.

\bibitem[Ebach \& Goujet, 2006]{Ebach2006}
Ebach, M.~C. \& Goujet, D.~F. (2006)\removeperiod.
\newblock The first biogeographical map.
\newblock {\em Journal of Biogeography}, \textbf{33}, 761--769.

\bibitem[Fenu \textit{et~al.}, 2014]{Fenu2014}
Fenu, G., Fois, M., Ca{\~n}adas, E.~M. \& Bacchetta, G. (2014)\removeperiod.
\newblock Using endemic-plant distribution, geology and geomorphology in
  biogeography: the case of {Sardinia} ({Mediterranean} {Basin}).
\newblock {\em Systematics and Biodiversity}, \textbf{12}, 181--193.

\bibitem[Ficetola \textit{et~al.}, 2017]{Ficetola2017}
Ficetola, G.~F., Mazel, F. \& Thuiller, W. (2017)\removeperiod.
\newblock Global determinants of zoogeographical boundaries.
\newblock {\em Nature Ecology \& Evolution}, \textbf{1}, 0089.

\bibitem[Flahault \& Durand, 1887]{Flahault1887}
Flahault, C. \& Durand, M. (1887)\removeperiod.
\newblock Limite de la r{\'e}gion m{\'e}diterran{\'e}enne en {France}.
\newblock {\em Publications de la Soci{\'e}t{\'e} Linn{\'e}enne de Lyon},
  \textbf{5}, 9--9.

\bibitem[Funk \textit{et~al.}, 2002]{Funk2002}
Funk, V.~A., Richardson, K.~S. \& Sakai, A.~K. (2002)\removeperiod.
\newblock Systematic {Data} in {Biodiversity} {Studies}: {Use} {It} or {Lose}
  {It}.
\newblock {\em Systematic Biology}, \textbf{51}, 303--316.

\bibitem[Garc{\'i}a-Barros \textit{et~al.}, 2002]{Garcia2002}
Garc{\'i}a-Barros, E., Gurrea, P., Luci{\'a}{\~n}ez, M.~J., Cano, J.~M.,
  Munguira, M.~L., Moreno, J.~C., Sainz, H., Sanz, M.~J. \& Sim{\'o}n, J.~C.
  (2002)\removeperiod.
\newblock Parsimony analysis of endemicity and its application to animal and
  plant geographical distributions in the {Ibero}-{Balearic} region (western
  {Mediterranean}).
\newblock {\em Journal of Biogeography}, \textbf{29}, 109--124.

\bibitem[Gaultier, 1989]{Gaultier1989}
Gaultier, C. (1989)\removeperiod.
\newblock {\em Relations entre pelouses eurosib{\'e}riennes (Festuco-Brometea
  {B}r. -{B}l. Et {T}x. 43) et groupements m{\'e}diterran{\'e}ens
  ({O}nonido-{R}osmarinetea {B}r. -{B}l. 47) : {\'e}tude r{\'e}gionale (Diois)
  et synth{\'e}se sur le pourtour m{\'e}diterran{\'e}en {N}ord-occidental}.
\newblock PhD thesis, Universit{\'e} de {P}aris-{S}ud.

\bibitem[Graham \& Hijmans, 2006]{Graham2006}
Graham, C.~H. \& Hijmans, R.~J. (2006)\removeperiod.
\newblock A comparison of methods for mapping species ranges and species
  richness.
\newblock {\em Global Ecology and Biogeography}, \textbf{15}, 578--587.

\bibitem[Greuter, 1991]{Greuter1991}
Greuter, W. (1991)\removeperiod.
\newblock {B}otanical diversity, endemism, rarity and extinction in the
  {M}editerranean area: an analysis based on the published volumes of
  {Med–Checklist}.
\newblock {\em Botanika Chronika}, \textbf{10}, 63--79.

\bibitem[Guimer{\`a} \& Nunes~Amaral, 2005]{Guimera2005}
Guimer{\`a}, R. \& Nunes~Amaral, L.~A. (2005)\removeperiod.
\newblock Functional cartography of complex metabolic networks.
\newblock {\em Nature}, \textbf{433}, 895--900.

\bibitem[Hampe \& Petit, 2005]{Hampe2005}
Hampe, A. \& Petit, R.~J. (2005)\removeperiod.
\newblock Conserving biodiversity under climate change: the rear edge matters.
\newblock {\em Ecology Letters}, \textbf{8}, 461--467.

\bibitem[Hewitt, 2000]{Hewitt2000}
Hewitt, G. (2000)\removeperiod.
\newblock The genetic legacy of the {Quaternary} ice ages.
\newblock {\em Nature}, \textbf{405}, 907--913.

\bibitem[Holt \textit{et~al.}, 2013]{Holt2013}
Holt, B.~G., Lessard, J.-P., Borregaard, M.~K., Fritz, S.~A., Ara{\'u}o, M.~B.,
  Dimitrov, D.and~Fabre, P.-H., Graham, C.~H., Graves, G.~R., J{\o}nsson,
  K.~A., Nogu{\'e}s-Bravo, D., Wang, Z., Whittaker, R.~J., Fjelds{\~a}, J. \&
  Rahbek, C. (2013)\removeperiod.
\newblock An {Update} of {Wallace}'s {Zoogeographic} {Regions} of the {World}.
\newblock {\em Science}, \textbf{339}.

\bibitem[Joly \textit{et~al.}, 2010]{Joly2010}
Joly, D., Brossard, T., Cardot, H., Cavailhes, J., Hilal, M. \& Wavresky, P.
  (2010)\removeperiod.
\newblock Les types de climats en {France}, une construction spatiale.
\newblock {\em Cybergeo : European Journal of Geography}.

\bibitem[Koleff \textit{et~al.}, 2003]{Koleff2003}
Koleff, P., Gaston, J. \& Lennon, J. (2003)\removeperiod.
\newblock Measuring beta diversity for presence-absence data.
\newblock {\em Journal of Animal Ecology}, \textbf{72}, 367--382.

\bibitem[Kougioumoutzis \textit{et~al.}, 2014]{Kougioumoutzis2014}
Kougioumoutzis, K., Simaiakis, S.~M. \& Tiniakou, A. (2014)\removeperiod.
\newblock Network biogeographical analysis of the central {{Aegean}}
  archipelago.
\newblock {\em Journal of Biogeography}, \textbf{41}, 1848--1858.

\bibitem[Kreft \& Jetz, 2010]{Kreft2010}
Kreft, H. \& Jetz, W. (2010)\removeperiod.
\newblock A framework for delineating biogeographical regions based on species
  distributions.
\newblock {\em Journal of Biogeography}, \textbf{37}, 2029--2053.

\bibitem[Lamarck \& Candolle, 1805]{Lamarck1805}
Lamarck, J. d. M.~d. \& Candolle, A. (1805)\removeperiod.
\newblock {\em Flore fran{\c c}aise, ou descriptions succinctes de toutes les
  plantes qui croissent naturellement en France, dispos{\'e}es selon une
  nouvelle m{\'e}thode d'analyse, et pr{\'e}c{\'e}d{\'e}es par un expos{\'e}
  des principes {\'e}l{\'e}mentaires de la botanique}.
\newblock Paris.

\bibitem[Lancichinetti \textit{et~al.}, 2010]{Lancichinetti2010}
Lancichinetti, A., Radicchi, F. \& Ramasco, J.~J. (2010)\removeperiod.
\newblock Statistical significance of communities in networks.
\newblock {\em Physical Review E}, \textbf{81}, 046110.

\bibitem[Lancichinetti \textit{et~al.}, 2011]{Lancichinetti2011}
Lancichinetti, A., Radicchi, F., Ramasco, J.~J. \& Fortunato, S.
  (2011)\removeperiod.
\newblock Finding {{Statistically Significant Communities}} in {{Networks}}.
\newblock {\em PLOS ONE}, \textbf{6}, e18961.

\bibitem[Latini \textit{et~al.}, 2017]{Latini2017}
Latini, M., Bartolucci, F., Conti, F., Iberite, M., Nicolella, G., Scoppola, A.
  \& Abbate, G. (2017)\removeperiod.
\newblock Detecting {Phytogeographic} {Units} {Based} on {Native} {Woody}
  {Flora}: {A} {Case} {Study} in {Central} {Peninsular} {Italy}.
\newblock {\em The Botanical Review}, \textbf{83}, 253--281.

\bibitem[Lavagne, 2008]{Lavagne2008}
Lavagne, A. (2008)\removeperiod.
\newblock Synth{\`e}se phytog{\'e}ographique du d{\'e}partement du {Var}.
\newblock In {\em Le Var et sa flore. Plantes rares ou prot{\'e}g{\'e}es.}
  Naturalia Publications, Inflovar.

\bibitem[Lavergne \textit{et~al.}, 2006]{Lavergne2006}
Lavergne, S., Molina, J. \& Debussche, M. (2006)\removeperiod.
\newblock Fingerprints of environmental change on the rare mediterranean flora:
  a 115-year study.
\newblock {\em Global Change Biology}, \textbf{12}, 1466--1478.

\bibitem[Lavergne \textit{et~al.}, 2004]{Lavergne2004}
Lavergne, S., Thompson, J.~D., Garnier, E. \& Debussche, M.
  (2004)\removeperiod.
\newblock The biology and ecology of narrow endemic and widespread plants: a
  comparative study of trait variation in 20 congeneric pairs.
\newblock {\em Oikos}, \textbf{107}, 505--518.

\bibitem[Lebart \textit{et~al.}, 2000]{Lebart2000}
Lebart, L., Piron, M. \& Morineau, A. (2000)\removeperiod.
\newblock {\em {Statistique exploratoire multidimensionnelle}}.
\newblock {Dunod}, Paris.

\bibitem[Lennon \textit{et~al.}, 2001]{Lennon2001}
Lennon, J.~J., Koleff, P., Greenwood, J. J.~D. \& Gaston, K.~J.
  (2001)\removeperiod.
\newblock The geographical structure of {{British}} bird distributions:
  Diversity, spatial turnover and scale.
\newblock {\em Journal of Animal Ecology}, \textbf{70}, 966--979.

\bibitem[Lled{\'o} \textit{et~al.}, 2005]{Lledo2005}
Lled{\'o}, M.~D., Crespo, M.~B., Fay, M.~F. \& Chase, M.~W.
  (2005)\removeperiod.
\newblock Molecular phylogenetics of {Limonium} and related genera
  ({Plumbaginaceae}): biogeographical and systematic implications.
\newblock {\em American Journal of Botany}, \textbf{92}, 1189--1198.

\bibitem[Lumaret \textit{et~al.}, 2002]{Lumaret2002}
Lumaret, R., Mir, C., Michaud, H. \& Raynal, V. (2002)\removeperiod.
\newblock Phylogeographical variation of chloroplast {DNA} in holm oak
  ({Quercus} ilex {L}.).
\newblock {\em Molecular Ecology}, \textbf{11}, 2327--2336.

\bibitem[M{\'e}dail \& Diadema, 2009]{Medail2009}
M{\'e}dail, F. \& Diadema, K. (2009)\removeperiod.
\newblock Glacial refugia influence plant diversity patterns in the
  {Mediterranean} {Basin}.
\newblock {\em Journal of Biogeography}, \textbf{36}, 1333--1345.

\bibitem[M{\'e}dail \& Qu{\'e}zel, 1997]{Medail1997}
M{\'e}dail, F. \& Qu{\'e}zel, P. (1997)\removeperiod.
\newblock Hot-{{Spots Analysis}} for {{Conservation}} of {{Plant Biodiversity}}
  in the {{Mediterranean Basin}}.
\newblock {\em Annals of the Missouri Botanical Garden}, \textbf{84}, 112--127.

\bibitem[Mikolajczak \textit{et~al.}, 2015]{Mikolajczak2015}
Mikolajczak, A., Maréchal, D., Sanz, T., Isenmann, M., Thierion, V. \& Luque,
  S. (2015)\removeperiod.
\newblock Modelling spatial distributions of alpine vegetation: {A} graph
  theory approach to delineate ecologically-consistent species assemblages.
\newblock {\em Ecological Informatics}, \textbf{30}, 196--202.

\bibitem[Molinier \& Tallon, 1970]{Molinier1970}
Molinier, R. \& Tallon, G. (1970)\removeperiod.
\newblock {\em Prodrome des unit{\'e}s phytosociologiques observ{\'e}es en
  {Camargue}}.
\newblock Molinier.

\bibitem[Molloy \& Reed, 1995]{Molloy1995}
Molloy, M. \& Reed, B. (1995)\removeperiod.
\newblock A critical point for random graphs with a given degree sequence.
\newblock {\em Random Structures \& Algorithms}, \textbf{6}, 161--180.

\bibitem[Morrone, 2018]{Morrone2018}
Morrone, J.~J. (2018)\removeperiod.
\newblock The spectre of biogeographical regionalization.
\newblock {\em Journal of Biogeography}, \textbf{105}, 1118--1123.

\bibitem[Mucina \textit{et~al.}, 2016]{Mucina2016}
Mucina, L., B{\"u}tmann, H., Dier{\ss}en, K., Theurillat, J.-P., Raus, T.,
  \v{C}arni, A., \v{S}umberov{\'a}, K., Willner, W., Dengler, J., Garc{\'i},
  R.~G., Chytr{\'y}, M., H{\'a}jek, M., Di~Pietro, R., Iakushenko, D., Pallas,
  J., Dani{\"e}ls, F.~J., Bergmeier, E., Santos~Guerra, A., Ermakov, N.,
  Valachovi\v{c}, M., Schamin{\'e}e, J. H.~J., Lysenko, T., Didukh, Y.~P.,
  Pignatti, S., Rodwell, J.~S., Capelo, J., Weber, H.~E., Solomeshch, A.,
  Dimopoulos, P., Aguiar, C., Hennekens, S.~M. \& Tich{\'y}, L.
  (2016)\removeperiod.
\newblock Vegetation of {Europe}: hierarchical floristic classification system
  of vascular plant, bryophyte, lichen, and algal communities.
\newblock {\em Applied Vegetation Science}, \textbf{19}, 3--264.

\bibitem[Murray, 1866]{Murray1866}
Murray, A. (1866)\removeperiod.
\newblock {\em The geographical distribution of mammals.}
\newblock Day and Son, limited,, London,.

\bibitem[Myers \textit{et~al.}, 2000]{Myers2000}
Myers, N., Mittermeier, R.~A., Mittermeier, C.~G., {da Fonseca}, G. A.~B. \&
  Kent, J. (2000)\removeperiod.
\newblock Biodiversity hotspots for conservation priorities.
\newblock {\em Nature}, \textbf{403}, 853--858.

\bibitem[Nieto~Feliner, 2014]{Nieto2014}
Nieto~Feliner, G. (2014)\removeperiod.
\newblock Patterns and processes in plant phylogeography in the {Mediterranean}
  {Basin}. {A} review.
\newblock {\em Perspectives in Plant Ecology, Evolution and Systematics},
  \textbf{16}, 265--278.

\bibitem[Ozenda, 1994]{Ozenda1994}
Ozenda, P. (1994)\removeperiod.
\newblock {\em V{\'e}g{\'e}tation du continent Europ{\'e}en}.
\newblock Delachaux et Niestl{\'e}, Lausanne.

\bibitem[Ozenda, 2002]{Ozenda2002}
Ozenda, P. (2002)\removeperiod.
\newblock {\em Perspectives pour une g{\'e}obiologie des montagnes}.
\newblock Presses Polytechniques et Universitaires Romandes, Lausanne.

\bibitem[Ozenda \& Lucas, 1987]{Ozenda1987}
Ozenda, P. \& Lucas, M.~J. (1987)\removeperiod.
\newblock Esquisse d'une carte de la v{\'e}g{\'e}tation potentielle de la
  france à 1/1 500 000.
\newblock {\em Documents de cartographie {\'e}cologique}.

\bibitem[Papuga \textit{et~al.}, 2018]{Papuga2018}
Papuga, G., Gauthier, P., Pons, V., Farris, E. \& Thompson, J.
  (2018)\removeperiod.
\newblock Ecological niche differentiation in peripheral populations: a
  comparative analysis of eleven mediterranean plant species.
\newblock {\em Ecography}, \textbf{176}, 724--738.

\bibitem[Papuga \textit{et~al.}, 2015]{Papuga2015}
Papuga, G., Gauthier, P., Ramos, J., Pons, V., Pironon, S., Farris, E. \&
  Thompson, J.~D. (2015)\removeperiod.
\newblock Range-{Wide} {Variation} in the {Ecological} {Niche} and {Floral}
  {Polymorphism} of the {Western} {Mediterranean} {Geophyte} {Narcissus} dubius
  {Gouan}.
\newblock {\em International Journal of Plant Sciences}, \textbf{176},
  724--738.

\bibitem[Pironon \textit{et~al.}, 2017]{Pironon2017}
Pironon, S., Papuga, G., Villellas, J., Angert, A.~L., Garc{\'i}a, M.~B. \&
  Thompson, J.~D. (2017)\removeperiod.
\newblock Geographic variation in genetic and demographic performance: new
  insights from an old biogeographical paradigm.
\newblock {\em Biological Reviews}, \textbf{92}, 1877--1909.

\bibitem[Pu{\c s}ca{\c s} \& Choler, 2012]{Puscas2012}
Pu{\c s}ca{\c s}, M. \& Choler, P. (2012)\removeperiod.
\newblock A biogeographic delineation of the {European} {Alpine} {System} based
  on a cluster analysis of {Carex} curvula-dominated grasslands.
\newblock {\em Flora - Morphology, Distribution, Functional Ecology of Plants},
  \textbf{207}, 168--178.

\bibitem[Qu{\'e}zel, 1999]{Quezel1999}
Qu{\'e}zel, P. (1999)\removeperiod.
\newblock Les grandes structures de v{\'e}g{\'e}tation en r{\'e}gion
  m{\'e}diterran{\'e}enne: {Facteurs} d{\'e}terminants dans leur mise en place
  post-glaciaire.
\newblock {\em Geobios}, \textbf{32}, 19--32.

\bibitem[Qu{\'e}zel \& M{\'e}dail, 2004]{Quezel2004}
Qu{\'e}zel, P. \& M{\'e}dail, F. (2004)\removeperiod.
\newblock {\em Ecologie et biog{\'e}ographie des for{\^e}ts du bassin
  m{\'e}dit{\'e}rran{\'e}en}.
\newblock Elsevier Masson, Paris.

\bibitem[Ricklefs, 1987]{Ricklefs1987}
Ricklefs, R.~E. (1987)\removeperiod.
\newblock Community diversity: relative roles of local and regional processes.
\newblock {\em Science (New York, N.Y.)}, \textbf{235}, 167--171.

\bibitem[Ricklefs, 2004]{Ricklefs2004}
Ricklefs, R.~E. (2004)\removeperiod.
\newblock A comprehensive framework for global patterns in biodiversity.
\newblock {\em Ecology Letters}, \textbf{7}, 1--15.

\bibitem[Rivas-Mart{\'i}nez \textit{et~al.}, 2002]{Rivas2002}
Rivas-Mart{\'i}nez, S., D{\'i}az, T., Fern{\'a}ndez-Gonz{\'a}lez, F., Izco, J.,
  Loidi, J., Lous{\~a}, M. \& Penas, A. (2002)\removeperiod.
\newblock Vascular plant communities of {Spain} and {Portugal}: addenda to the
  syntaxonomical checklist of 2001. {Part} {II}.
\newblock {\em Itinera Geobot.}, \textbf{15(2)}, 433--922.

\bibitem[Rivas-Mart{\'i}nez \textit{et~al.}, 2004a]{Rivas2004a}
Rivas-Mart{\'i}nez, S., Penas, A. \& D{\'i}az, T. (2004a)\removeperiod.
\newblock {\em Bioclimatic {Map} of {Europe}}.
\newblock University of Le{\'o}n.

\bibitem[Rivas-Mart{\'i}nez \textit{et~al.}, 2004b]{Rivas2004}
Rivas-Mart{\'i}nez, S., Penas, A. \& D{\'i}az, T. (2004b)\removeperiod.
\newblock {\em Biogeographic {Map} of {Europe}}.
\newblock University of Le{\'o}n.

\bibitem[Rosenbaum \textit{et~al.}, 2002]{Rosenbaum2002}
Rosenbaum, G., Lister, G.~S. \& Duboz, C. (2002)\removeperiod.
\newblock Reconstruction of the tectonic evolution of the western
  {Mediterranean} since the {Oligocene}.
\newblock {\em Journal of the Virtual Explorer}, \textbf{8}, 107--130.

\bibitem[Rosvall \& Bergstrom, 2008]{Rosvall2008}
Rosvall, M. \& Bergstrom, C.~T. (2008)\removeperiod.
\newblock Maps of random walks on complex networks reveal community structure.
\newblock {\em Proceedings of the National Academy of Sciences}, \textbf{105},
  1118--1123.

\bibitem[Rousseeuw, 1987]{Rousseeuw1987}
Rousseeuw, P.~J. (1987)\removeperiod.
\newblock Silhouettes: A graphical aid to the interpretation and validation of
  cluster analysis.
\newblock {\em Journal of Computational and Applied Mathematics}, \textbf{20},
  53 -- 65.

\bibitem[Rundel \textit{et~al.}, 2016]{Rundel2016}
Rundel, P.~W., Arroyo, M., Cowling, R.~M., Keeley, J.~E., Lamont, B.~B. \&
  Vargas, P. (2016)\removeperiod.
\newblock Mediterranean {Biomes}: {Evolution} of {Their} {Vegetation},
  {Floras}, and {Climate}.
\newblock {\em Annual Review of Ecology, Evolution, and Systematics},
  \textbf{47}, 383--407.

\bibitem[Rushton \textit{et~al.}, 2004]{Rushton2004}
Rushton, S.~P., Ormerod, S.~J. \& Kerby, G. (2004)\removeperiod.
\newblock New paradigms for modelling species distributions?
\newblock {\em Journal of Applied Ecology}, \textbf{41}, 193--200.

\bibitem[Saiz \textit{et~al.}, 1998]{Saiz1998}
Saiz, J. C.~M., Parga, I.~C. \& Ollero, H.~S. (1998)\removeperiod.
\newblock Numerical analyses of distributions of {Iberian} and {Balearic}
  endemic monocotyledons.
\newblock {\em Journal of Biogeography}, \textbf{25}, 179--194.

\bibitem[Stewart \& Lister, 2001]{Stewart2001}
Stewart, J.~R. \& Lister, A.~M. (2001)\removeperiod.
\newblock Cryptic northern refugia and the origins of the modern biota.
\newblock {\em Trends in Ecology \& Evolution}, \textbf{16}, 608--613.

\bibitem[Stoddart, 1992]{Stoddart1992}
Stoddart, D.~R. (1992)\removeperiod.
\newblock {\em Biogeography of the {{Tropical Pacific}}}.
\newblock University of Hawaii Press.

\bibitem[Tassin, 2017]{Tassin2017}
Tassin, C. (2017)\removeperiod.
\newblock {\em Paysages v{\'e}g{\'e}taux du domaine m{\'e}diterran{\'e}en :
  {Bassin} m{\'e}diterran{\'e}en, {Californie}, {Chili} central, {Afrique} du
  {Sud}, {Australie} m{\'e}ridionale}.
\newblock Référence. IRD Éditions, Marseille.

\bibitem[Thompson, 2005]{Thompson2005a}
Thompson, J.~D. (2005)\removeperiod.
\newblock {\em Plant {Evolution} in the {Mediterranean}}.
\newblock Oxford University Press, Oxford, New York.

\bibitem[Thompson \textit{et~al.}, 2005]{Thompson2005b}
Thompson, J.~D., Lavergne, S., Affre, L., Gaudeul, M. \& Debussche, M.
  (2005)\removeperiod.
\newblock Ecological {Differentiation} of {Mediterranean} {Endemic} {Plants}.
\newblock {\em Taxon}, \textbf{54}, 967--976.

\bibitem[Tison \& Foucault, 2014]{Tison2014}
Tison, J.-M. \& Foucault, B.~d. (2014)\removeperiod.
\newblock {\em Flora {Gallica} : {Flore} de {France}}.
\newblock Biotope Editions, Mèze.

\bibitem[Tison \textit{et~al.}, 2014]{Tison2014a}
Tison, J.-M., Jauzein, P. \& Michaud, H. (2014)\removeperiod.
\newblock {\em Flore de la {France} m{\'e}diterranéenne continentale}.
\newblock Naturalia Publications, Turriers, 1ère édition edition.

\bibitem[Turner \textit{et~al.}, 2001]{Turner2001}
Turner, M.~G., Gardner, R.~H. \& O'Neill, R.~V. (2001)\removeperiod.
\newblock {\em Landscape {Ecology} in {Theory} and {Practice}: {Pattern} and
  {Process}}.
\newblock Springer-Verlag.

\bibitem[Vilhena \& Antonelli, 2015]{Vilhena2015}
Vilhena, D.~A. \& Antonelli, A. (2015)\removeperiod.
\newblock A network approach for identifying and delimiting biogeographical
  regions.
\newblock {\em Nature Communications}, \textbf{6}, 6848.

\bibitem[Wahlenberg, 1812]{Wahlenberg1812}
Wahlenberg, G. (1812)\removeperiod.
\newblock {\em Flora {Lapponica}}.
\newblock Taberna libraria scholae realis, Berolini.

\bibitem[Wallace, 1876]{Wallace1876}
Wallace, A.~R. (1876)\removeperiod.
\newblock {\em The geographical distribution of animals : with a study of the
  relations of living and extinct faunas as elucidating the past changes of the
  earth's surface}.
\newblock Harper and Brothers, Publishers, New York.

\bibitem[Walter \& Breckle, 1991]{Walter1991}
Walter, H. \& Breckle, S.-W. (1991)\removeperiod.
\newblock {\em {\"O}kologie der Erde. Band 4: Spezielle {\"O}kologie der
  Gem{\"a}{\ss}igten und Arktischen Zonen au{\ss}erhalb Euro-Nordasiens}.
\newblock Gustav Fischer Verlag.

\bibitem[Walter \& Breckle, 1994]{Walter1994}
Walter, H. \& Breckle, S.-W. (1994)\removeperiod.
\newblock {\em {\"O}kologie der Erde. Band 3: Spezielle {\"O}kologie der
  Gem{\"a}{\ss}igten und Arktischen Zonen au{\ss}erhalb Euro-Nordasiens}.
\newblock Gustav Fischer Verlag.

\bibitem[Wilson \& Shmida, 1984]{Wilson1984}
Wilson, M.~V. \& Shmida, A. (1984)\removeperiod.
\newblock Measuring {{Beta Diversity}} with {{Presence}}-{{Absence Data}}.
\newblock {\em Journal of Ecology}, \textbf{72}, 1055--1064.

\end{thebibliography}

\onecolumngrid

\makeatletter
\renewcommand{\fnum@figure}{\sf\textbf{\figurename~\textbf{S}\textbf{\thefigure}}}
\renewcommand{\fnum@table}{\sf\textbf{\tablename~\textbf{S}\textbf{\thetable}}}
\makeatother

\setcounter{figure}{0}
\setcounter{table}{0}
\setcounter{equation}{0}

\vspace{1cm}
\section*{Appendix}

\subsection*{Interactive web application}

\begin{figure}[!ht]
	\centering 
	\includegraphics[width=13cm]{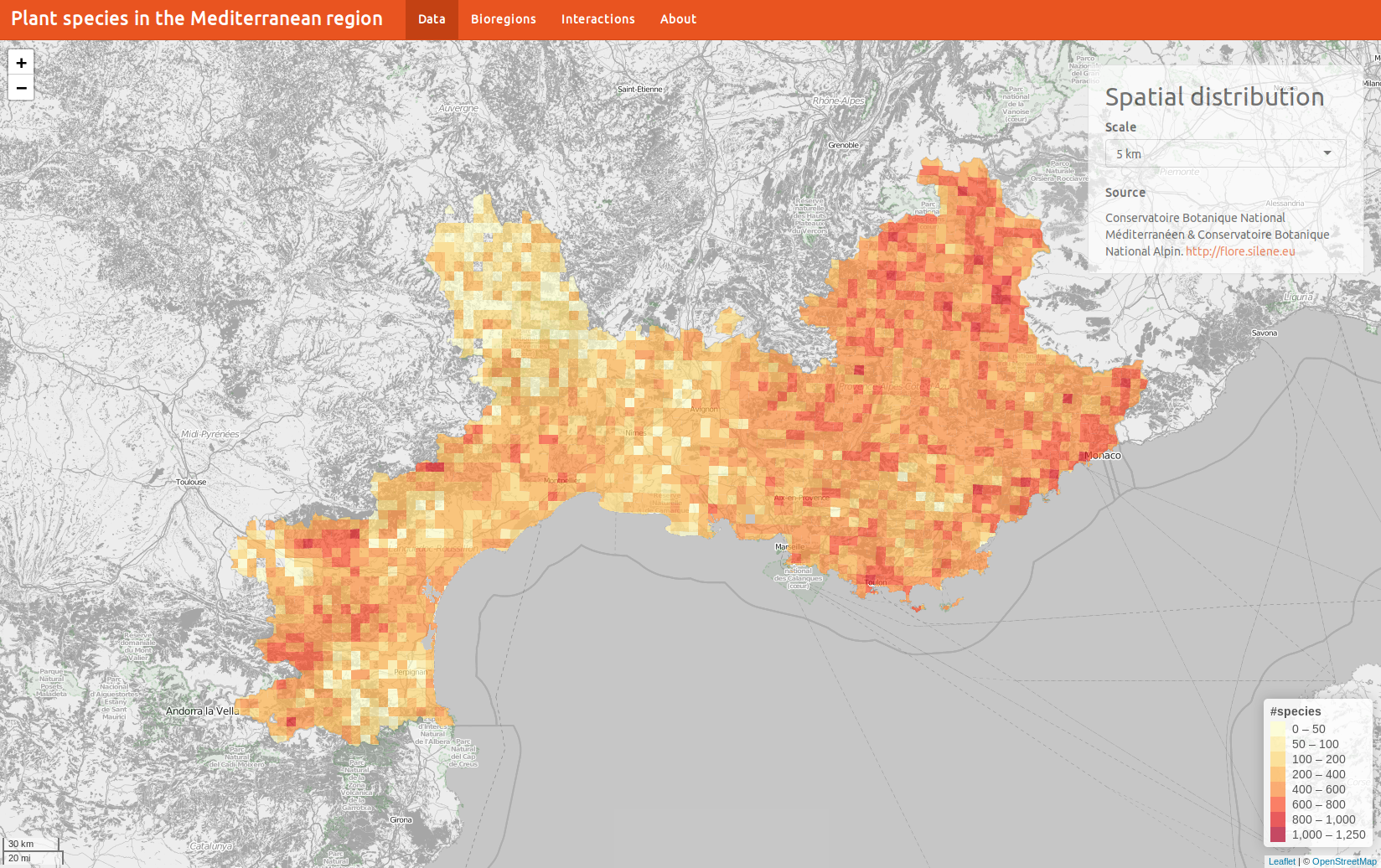}
	\caption{\textbf{Screenshot of the interactive web application.} \label{FigS0}}
\end{figure}

An interactive web application has been designed to provide an easy-to-use interface to visualize the results and the maps of the main paper and the Appendix (Figure S\ref{FigS0}). The source code of the interactive web application\footnote[1]{~ \url{https://maximelenormand.shinyapps.io/Biogeo/}} can be downloaded from\footnote[2]{~\url{www.maximelenormand.com/Codes}}.

\newpage
\clearpage
\newpage
\subsection*{Influence of scale on the biogeographical regions delineation}

\begin{figure}[!ht]
	\begin{center}
		\includegraphics[width=\linewidth]{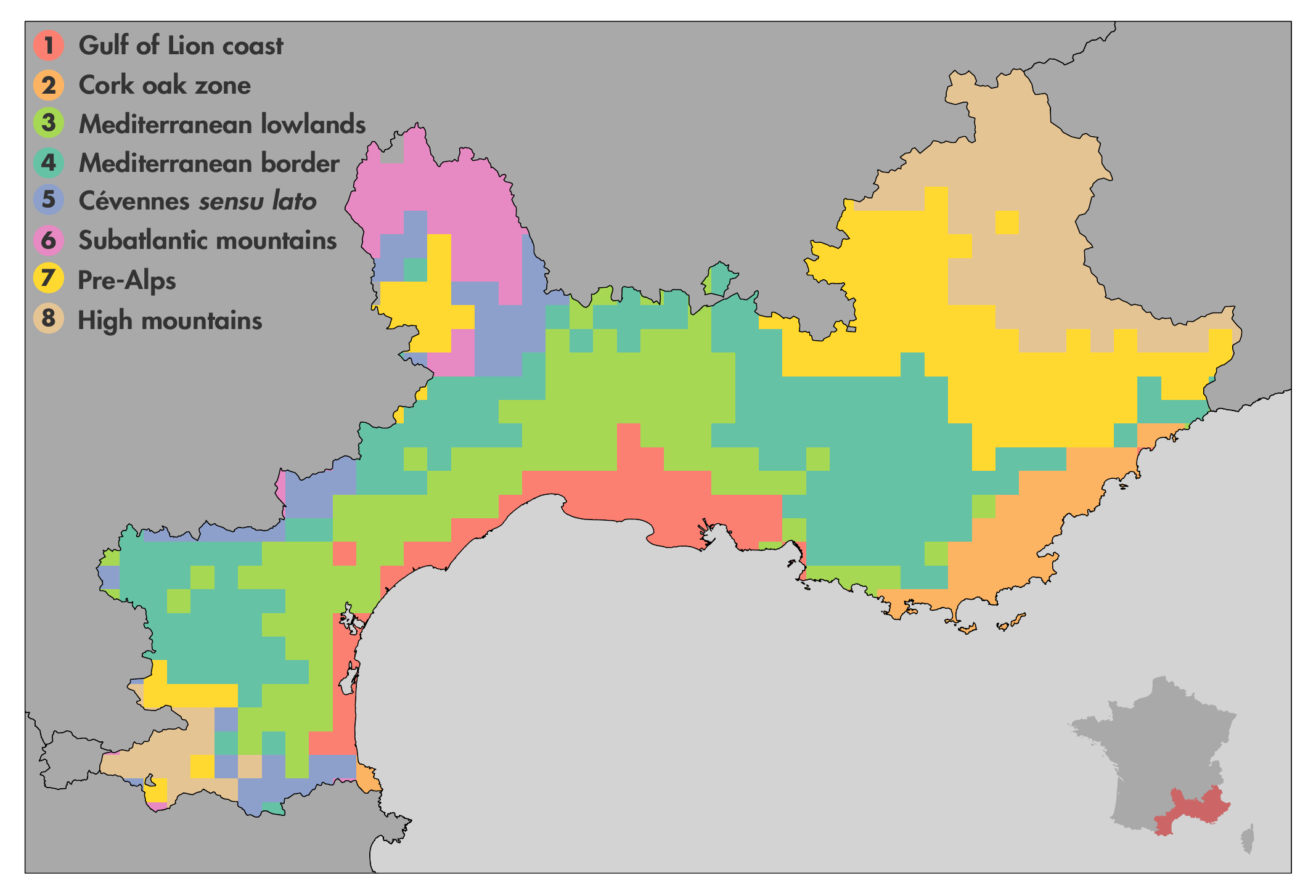}
		\caption{\sf \textbf{Biogeographical regions based on similarity in plant species (l = 10 km).} Eight biogeographical regions have been identified. 1. Gulf of Lion coast in red. 2. Cork oak zone in orange. 3. Mediterranean lowlands in light green. 4. Mediterranean border in dark green. 5. C{\'e}vennes \emph{sensu lato} in purple. 6. Subatlantic mountains in pink. 7. Prealps and other medium mountains in yellow. 8. High mountains in brown. \label{FigS1}}
	\end{center}
\end{figure}

In order to assess the impact of the spatial resolution on the results, we also applied the analysis with a grid composed of squares of lateral size $l=10$ km (Figure S\ref{FigS1}).
The spatial coherence, defined as the ratio between the number of grid cells in the largest patch and the total number of grid cells \citep{Turner2001}, is displayed for both scale in Table S\ref{TabS1}. 

\begin{table}[!h]
	\caption{\textbf{Spatial coherence of the biogeographical regions according to the scale.}}
	\label{TabS1}
	\begin{center}
		\begin{tabular}{|c|c|c|c|c|}			
			\hline
			\textbf{Bioregion} & \textbf{$n_i$ (l=5)} & \textbf{$SP_i$ (l=5)} & \textbf{$n_i$ (l=10)} & \textbf{$SP_i$ (l=10)} \\
			\hline
			
			\textbf{1} & 170 & 0.63 & 63  & 0.75 \\	
			\hline				
			\textbf{2} & 183 & 0.73 & 47  & 0.87 \\					
			\hline
			\textbf{3} & 529 & 0.86 & 124 & 0.83 \\
			\hline			
			\textbf{4} & 807 & 0.57 & 164 & 0.51 \\
			\hline						
			\textbf{5} & 120 & 0.50 & 45  & 0.31 \\
			\hline
			\textbf{6} & 152 & 0.70 & 48  & 0.79 \\
			\hline
			\textbf{7} & 400 & 0.67 & 114 & 0.75 \\
			\hline
			\textbf{8} & 246 & 0.78	& 110 & 0.77 \\
			\hline
			
		\end{tabular}
	\end{center}
\end{table}

\newpage
\clearpage
\newpage
\subsection*{Comparison of OSLOM to standard clustering methods}

In this work, bioregions are delineating using the community detection algorithm OSLOM applied on a weighted undirected spatial network whose intensity of links between grid cells are measured with the Jaccard similarity coefficient. This algorithm is nonparametric in the sense that it identifies statistically significant communities with respect to a global null model, and therefore the number of communities does not need to be defined \textit{a priori}. In order to assess the accuracy of the method, we compared the results obtained with OSLOM with the ones obtained with standard hierarchical clustering methods. Not that these standard methods cannot be directly applied on the spatial network described above, we first need to transform the network into a dissimilarity matrix. Three different agglomeration methods have been tested: average (UPGMA), mcquitty (WPGMA) and Ward\footnote{method="average", "mcquitty" and "ward.D2" with the hclust R function}. To choose the number of clusters, we used the average silhouette index $\bar{S}$ \citep{Rousseeuw1987}. For each cell $g$, we can compute $a(g)$ the average dissimilarity of $g$ (based on the Jaccard index in our case) with all the other cells in the cluster to which $g$ belongs. In the same way, we can compute the average dissimilarities of $g$ to the other clusters and define $b(g)$ as the lowest average dissimilarity among them. Using these two quantities, we compute the silhouette index $s(g)$ defined as,
\begin{equation}
s(g)=\frac{b(g)-a(g)}{max\{a(g),b(g)\}}
\label{sg}
\end{equation}
which measures how well clustered $g$ is. This measure is comprised between $-1$ for a very poor clustering quality and $1$ for an appropriately clustered $g$. We choose the number of clusters that maximize the average silhouette index over all the grid cells $\bar{S}=\sum_{g=1}^n s(g)/n$.

\begin{figure}[!ht]
	\begin{center}
		\includegraphics[width=10cm]{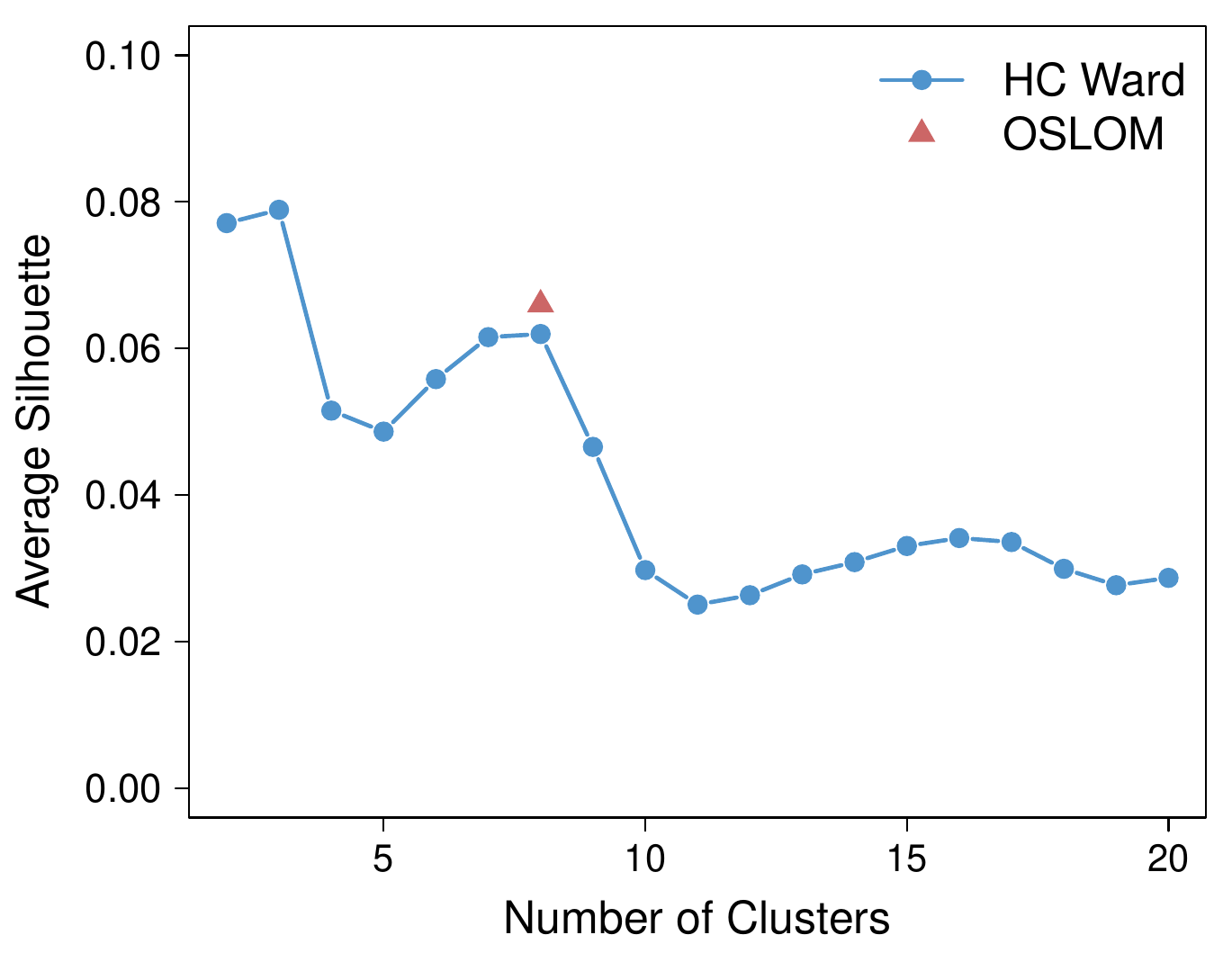}
		\caption{\sf \textbf{Average Silhouette as a function of the number of clusters obtained with Ward's clustering (in blue) and OSLOM (in red).} \label{FigS2}}
	\end{center}
\end{figure}

UPGMA and WPGMA failed to detect any coherent partitions, most of the grid cells were gathered in a giant cluster component even increasing significantly the number of clusters. Better results were obtained with Ward's method. The average Silhouette index as a function of the number of clusters is shown in Figure S\ref{FigS2}. 

Two optimal partitions have been detected with the average Silhouette index. It is interesting to note that the number of clusters of the second partition is the same that the one automatically detected with OSLOM (Figure S\ref{FigS2}).

\begin{figure}[!ht]
	\begin{center}
		\includegraphics[width=\linewidth]{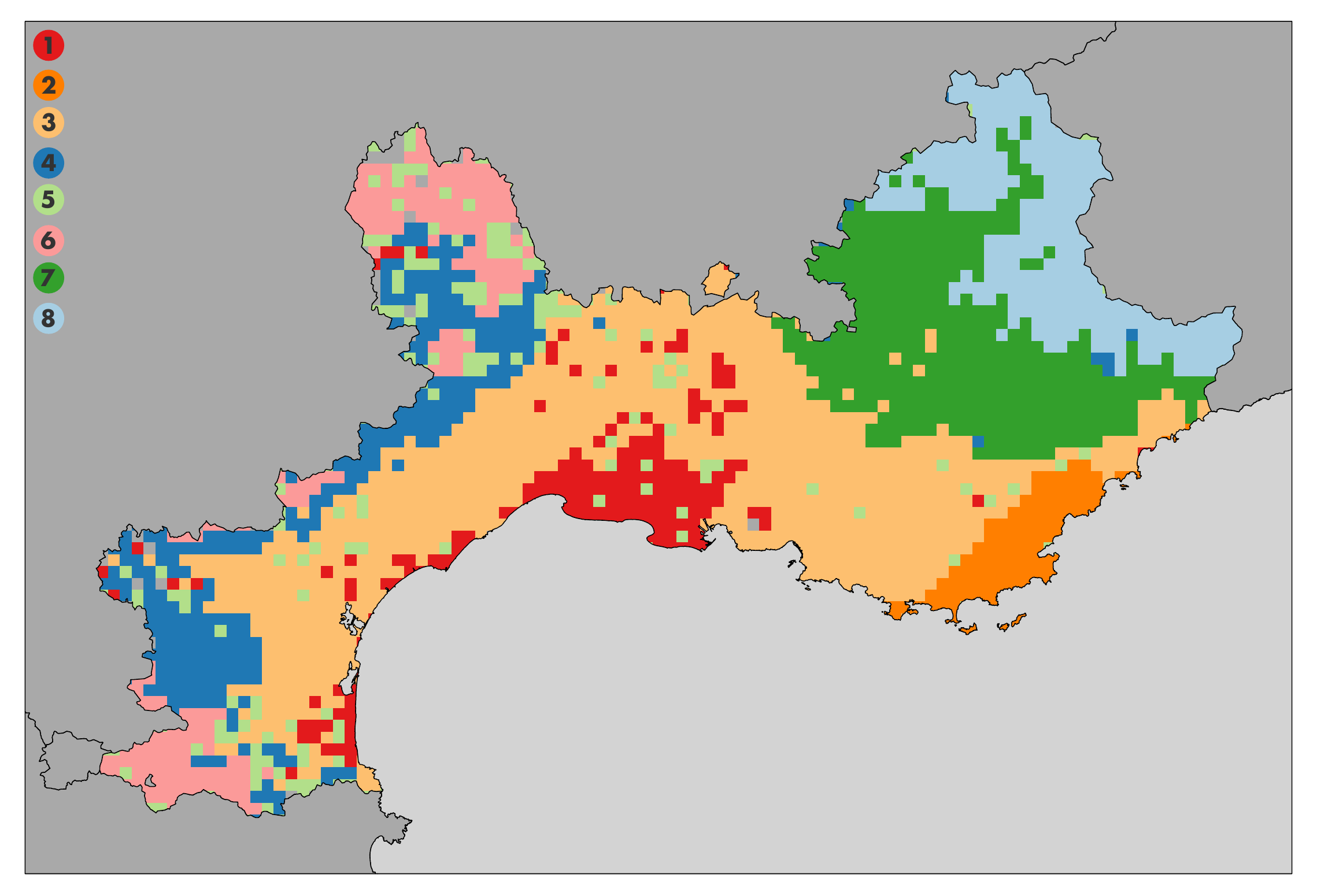}
		\caption{\sf \textbf{Biogeographical regions based on similarity in plant species obtained with Ward's clustering (l = 5 km).} Eight biogeographical regions have been identified. \label{FigS3}}
	\end{center}
\end{figure} 

A map of the eight optimal bioregions obtained with Ward's method is display in Figure S\ref{FigS3}. In order to compare the two partitions a contingency between the partitions obtained with Ward's method and OSLOM is shown in Table S\ref{TabS2}.

\begin{table}[!h]
	\caption{\textbf{Contingency tables between the partitions obtained with Ward (in row) and OSLOM (in column).}}
	\label{TabS2}
	\begin{center}
		\begin{tabular}{|c|c|c|c|c|c|c|c|c|}
			\hline
			\textbf{Bioregion} & \textbf{1} & \textbf{2} & \textbf{3} & \textbf{4} & \textbf{5} & \textbf{6} & \textbf{7} & \textbf{8} \\
			\hline
			
			\textbf{1} & \textbf{125} & 2 & 54	& 0	& 4	& 0	& 0	& 0 \\
			\hline
			\textbf{2} & 0 & \textbf{115} &	0 &	2 &	0 & 0 & 0 & 0 \\
			\hline
			\textbf{3} & 28	& 47 & \textbf{435} & 396 &	0 &	0 &	0 &	0 \\
			\hline
			\textbf{4} & 0 & 4 & 4 & \textbf{187} &	48 & 1 & 43 & 1 \\
			\hline
			\textbf{5} & 17	& 15 & 34 &	20 & \textbf{53} & 32 & 8 & 15 \\
			\hline
			\textbf{6} & 0 & 0 & 0 & 0 & 15 & \textbf{119} & 38 & 33 \\
			\hline
			\textbf{7} & 0	& 0 & 0 & 200 &	0 &	0 &	\textbf{216} & 0 \\
			\hline
			\textbf{8} & 0	& 0	& 0	& 2 & 0 & 0 & 95 & \textbf{197} \\
			\hline
			
		\end{tabular}
	\end{center}
\end{table}

\newpage
\clearpage
\newpage
\subsection*{Comparison with other delineations}

\begin{figure}[!h]
	\begin{center}
		\includegraphics[width=13cm]{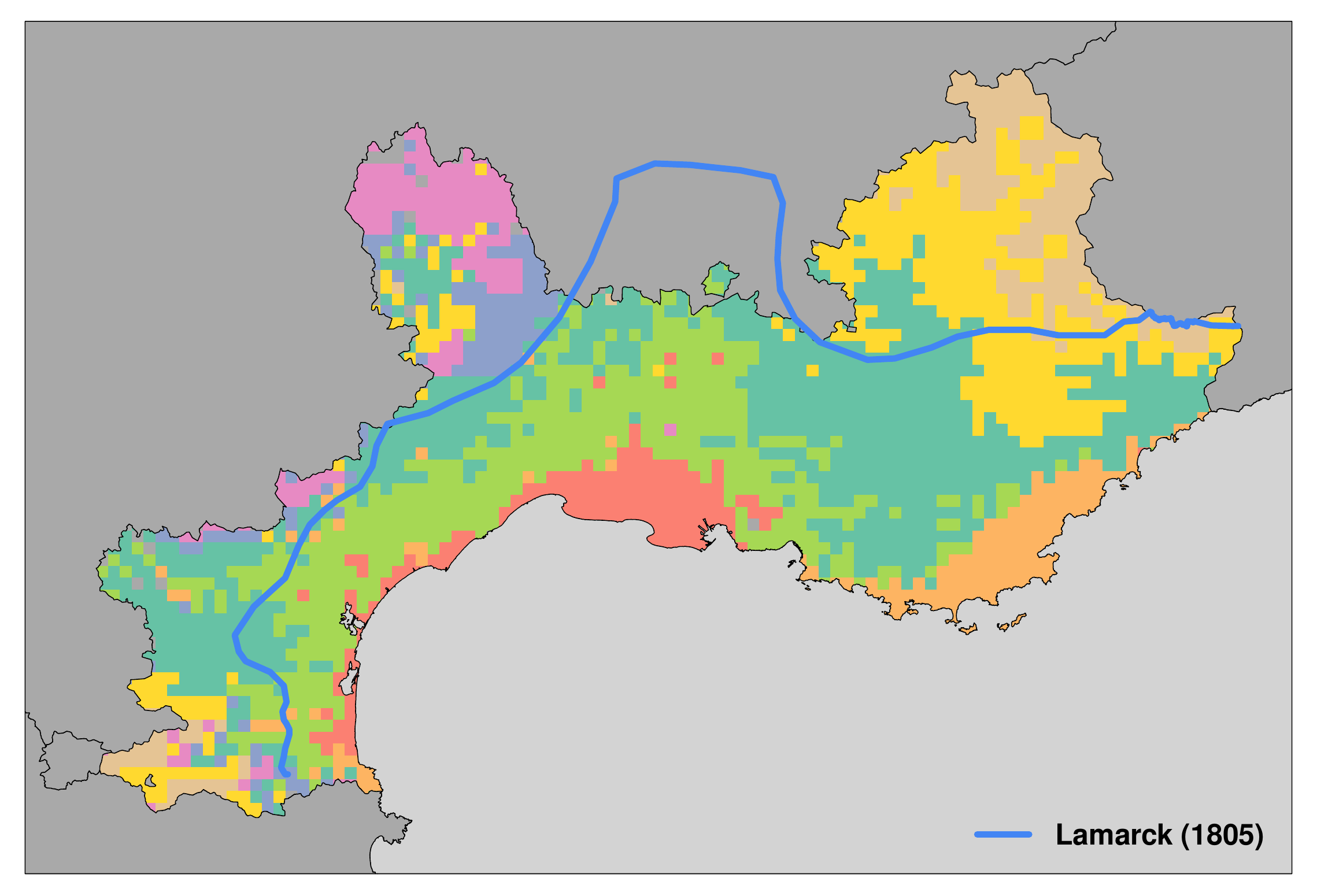}
		\caption{\sf \textbf{Comparison of the results obtained with OSLOM (l = 5 km) with Lamarck's limit of the Mediterranean level \citep{Lamarck1805,Ebach2006}.} \label{FigS4}}
	\end{center}
\end{figure} 

\begin{figure}[!h]
	\begin{center}
		\includegraphics[width=13cm]{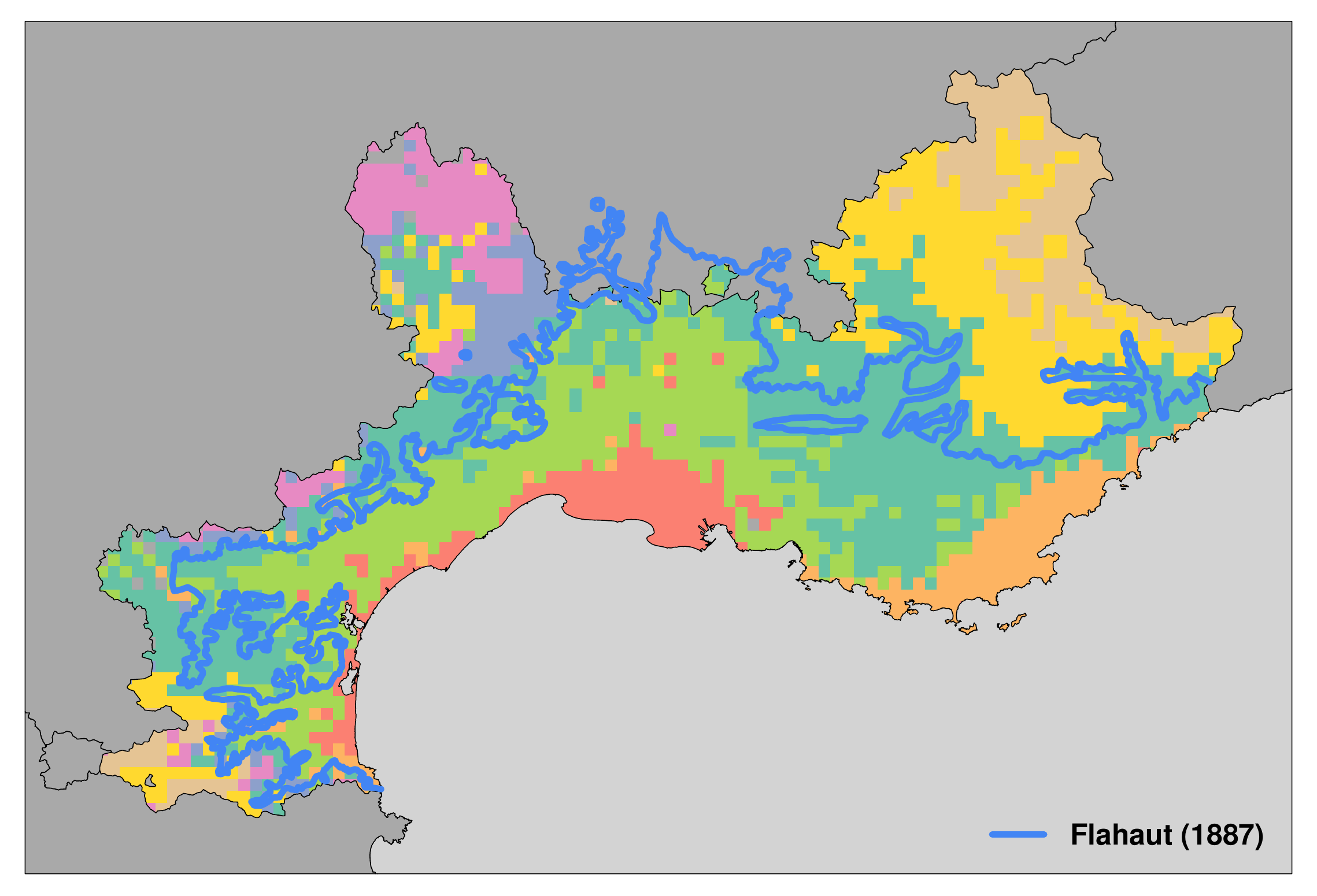}
		\caption{\sf \textbf{Comparison of the results obtained with OSLOM (l = 5 km) with Flahaut limit of the olive tree distribution \citep{Flahault1887}.}\label{FigS5}}
	\end{center}
\end{figure} 

\begin{figure}[!h]
	\begin{center}
		\includegraphics[width=13cm]{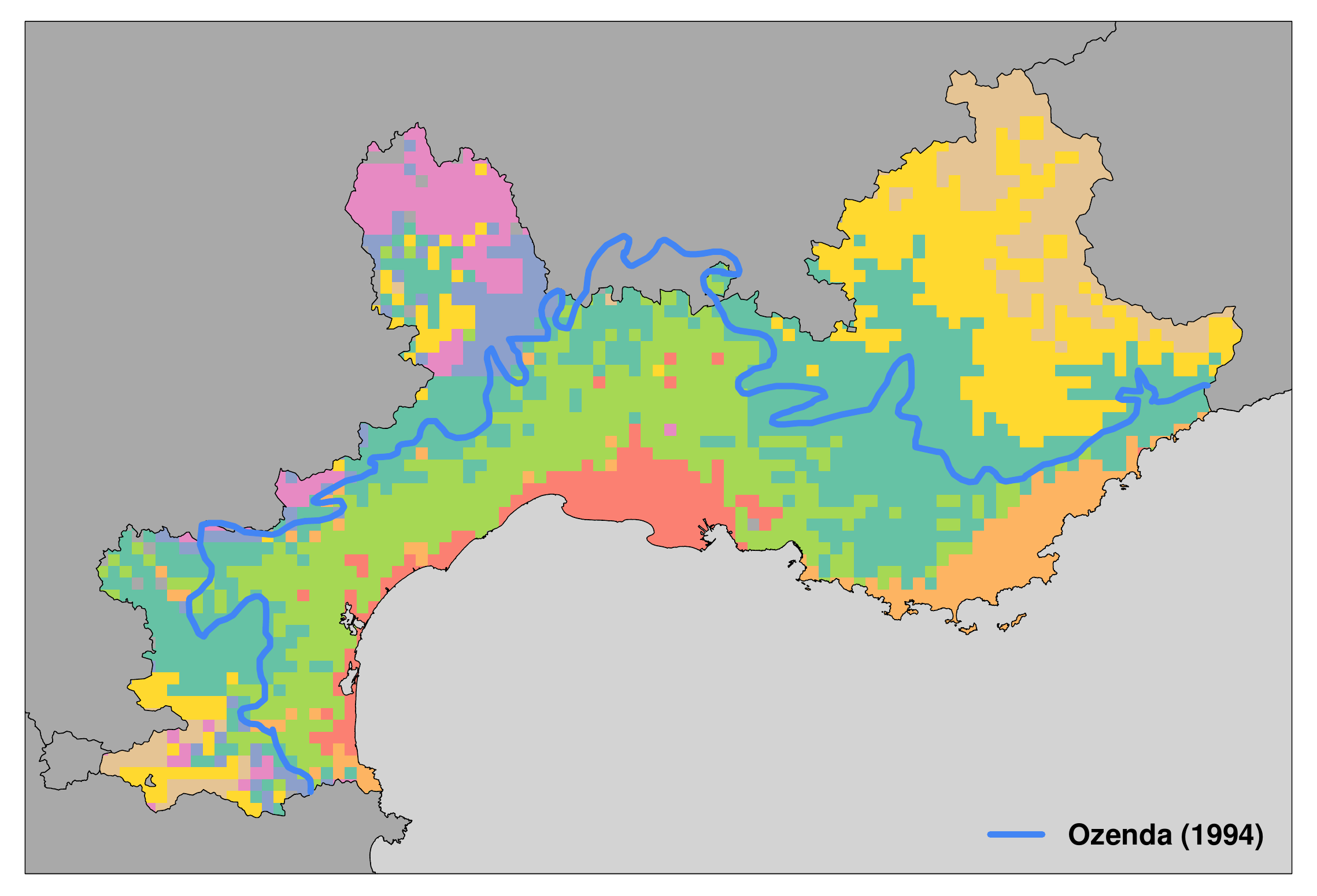}
		\caption{\sf \textbf{Comparison of the results obtained with OSLOM (l = 5 km) with Ozenda's mediterranean/supramediterranean limit \citep{Ozenda1994}.}\label{FigS6}}
	\end{center}
\end{figure} 

\begin{figure}[!h]
	\begin{center}
		\includegraphics[width=13.5cm]{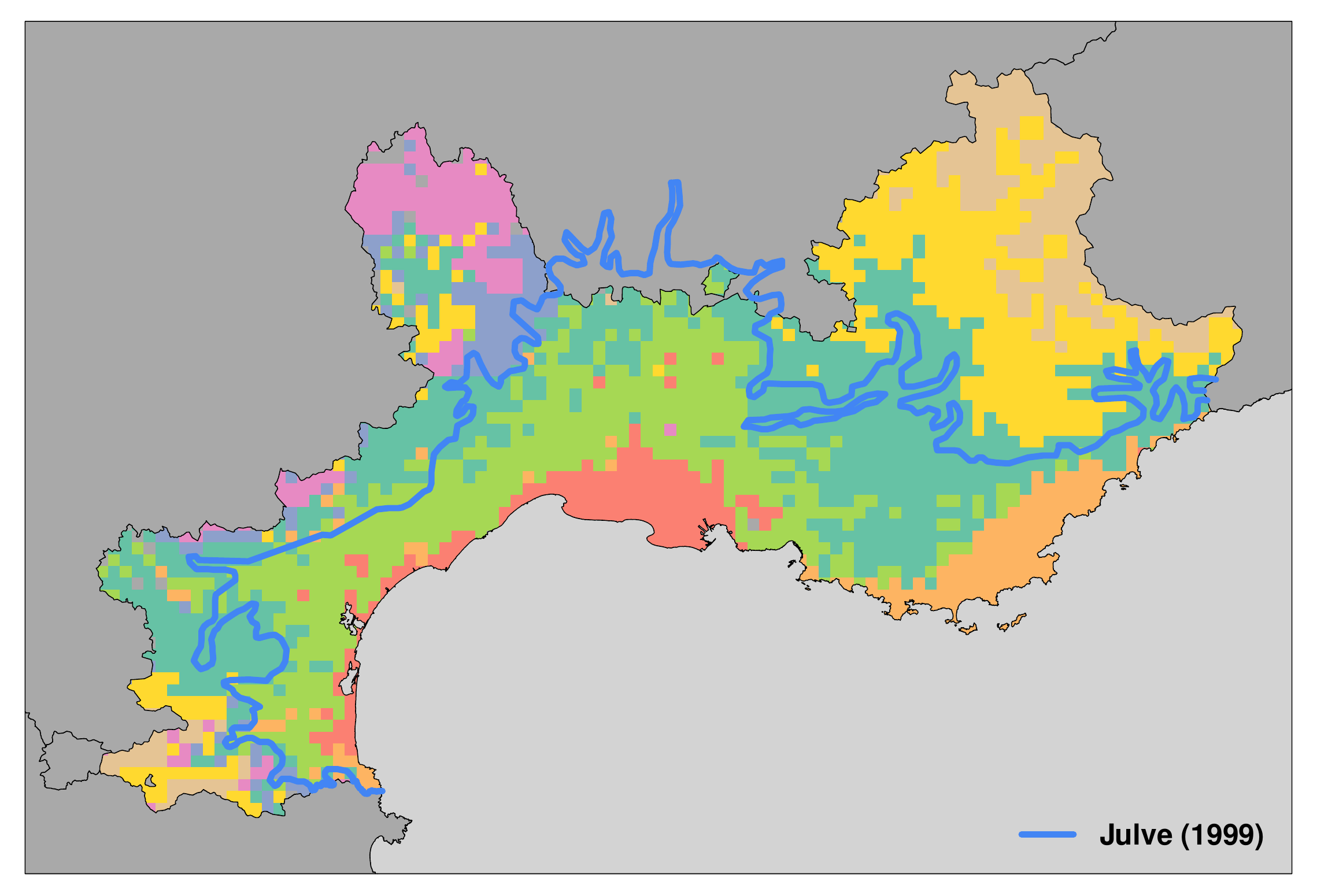}
		\caption{\sf \textbf{Comparison of the results obtained with OSLOM (l = 5 km) with Julve's mediterranean/supramediterranean limit \citep{Julve1999}.}\label{FigS7}}
	\end{center}
\end{figure} 

\begin{figure}[!h]
	\begin{center}
		\includegraphics[width=13.5cm]{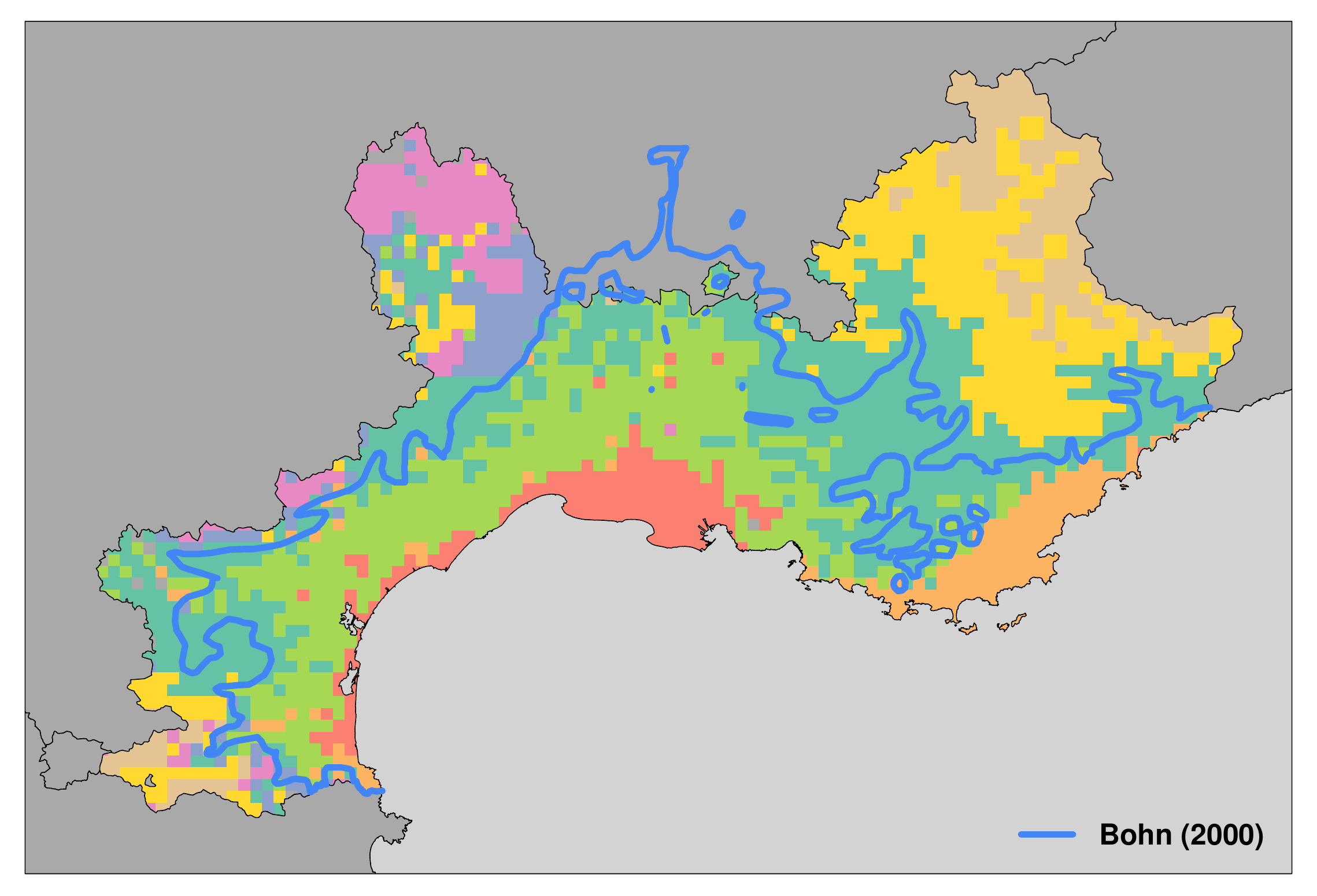}
		\caption{\sf \textbf{Comparison of the results obtained with OSLOM (l = 5 km) with Bohn's mediterranean/supramediterranean limit \citep{Bohn2000}.}\label{FigS8}}
	\end{center}
\end{figure} 

\begin{figure}[!h]
	\begin{center}
		\includegraphics[width=13.5cm]{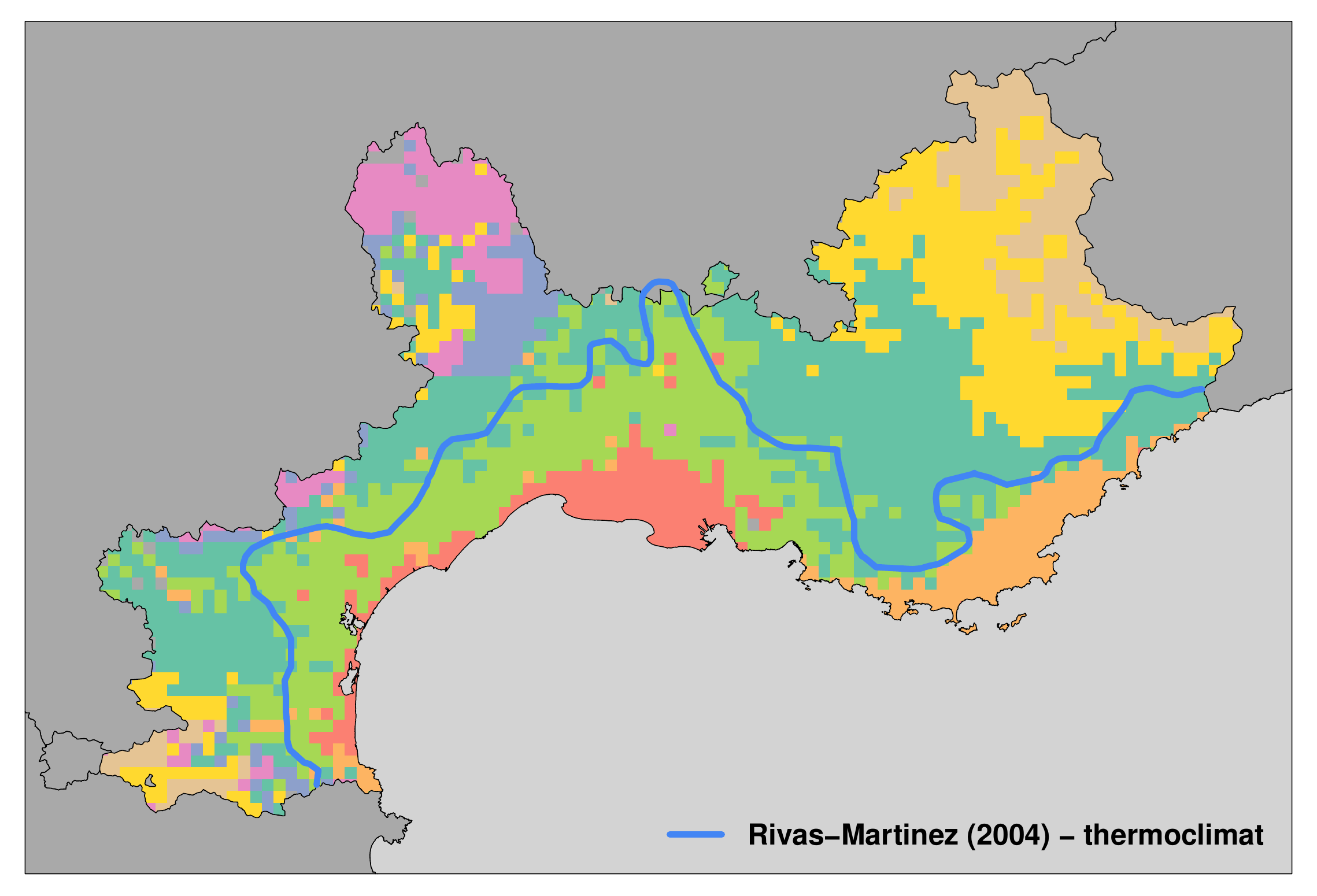}
		\caption{\sf \textbf{Comparison of the results obtained with OSLOM (l = 5 km) with Rivas-Mart{\'i}nez's thermoclimatic limit \citep{Rivas2004a}.}\label{FigS9}}
	\end{center}
\end{figure} 

\begin{figure}[!h]
	\begin{center}
		\includegraphics[width=13.5cm]{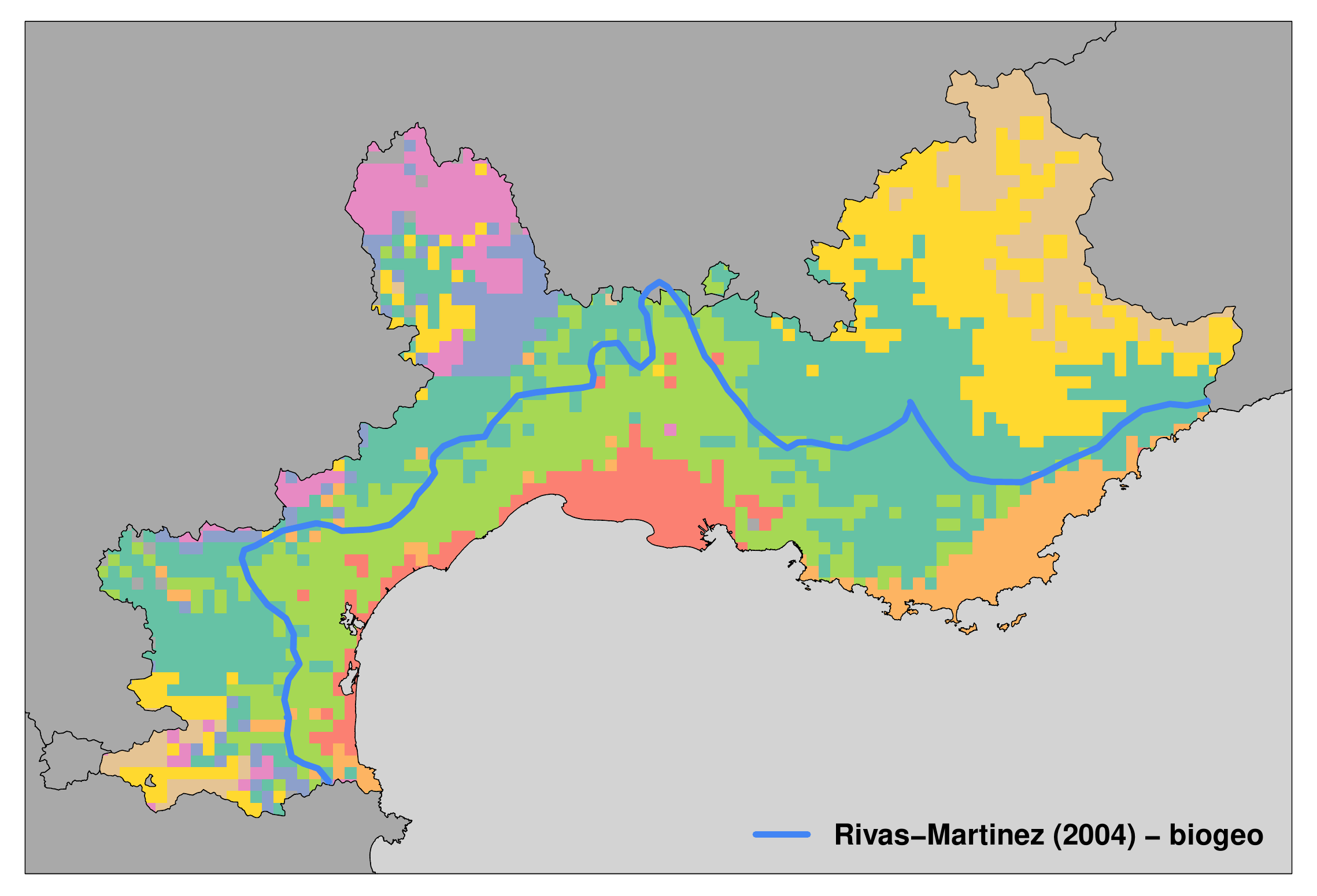}
		\caption{\sf \textbf{Comparison of the results obtained with OSLOM (l = 5 km) with Rivas-Mart{\'i}nez's biogeographical limit \citep{Rivas2004}.}\label{FigS10}}
	\end{center}
\end{figure} 

\begin{figure}[!h]
	\begin{center}
		\includegraphics[width=13.5cm]{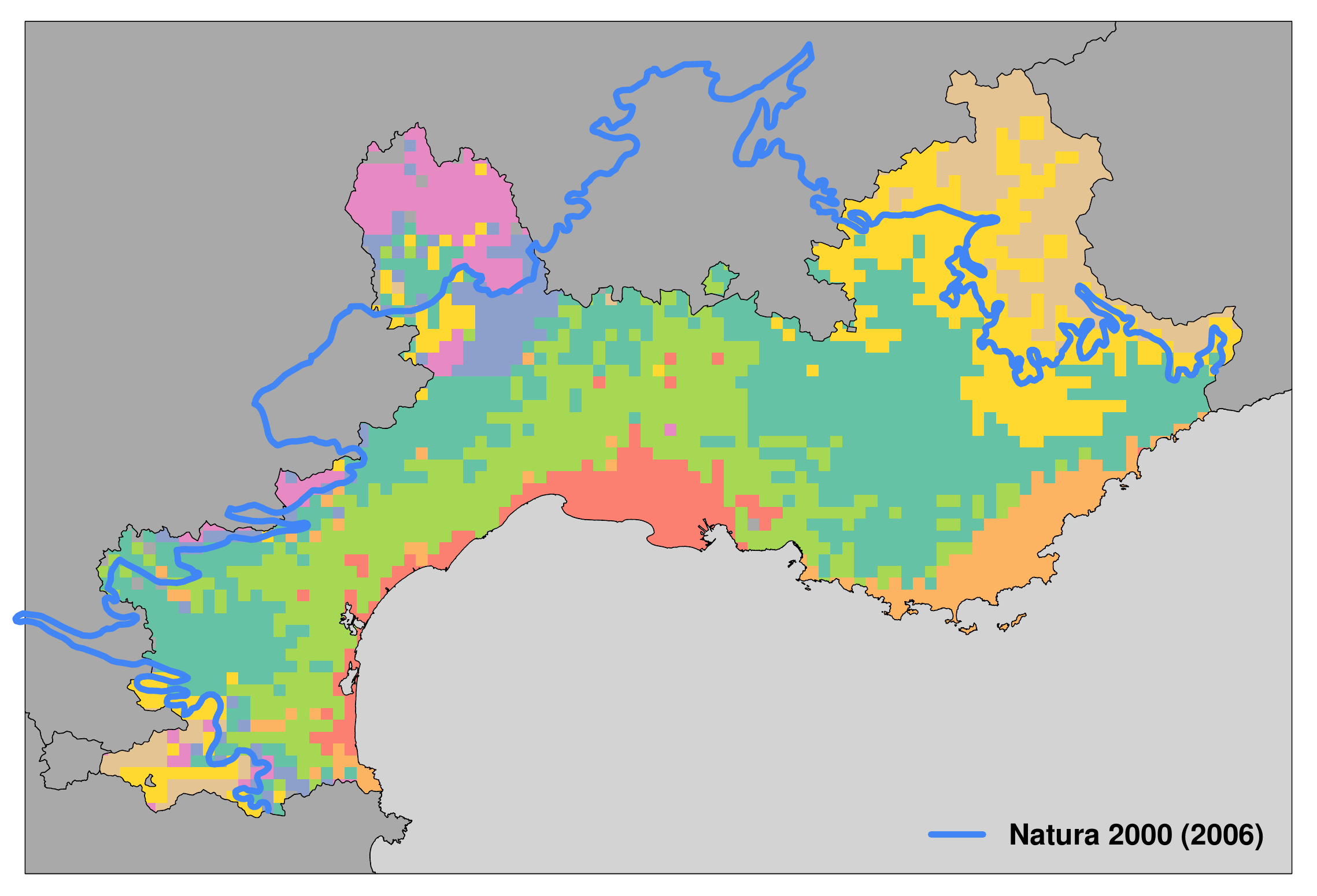}
		\caption{\sf \textbf{Comparison of the results obtained with OSLOM (l = 5 km) with the Natura 2000's limit \citep{EEA2006}.}\label{FigS11}}
	\end{center}
\end{figure} 

\newpage
\clearpage
\newpage
\subsection*{Supplementary Figures}

\begin{figure}[!h]
	\begin{center}
		\includegraphics[width=\linewidth]{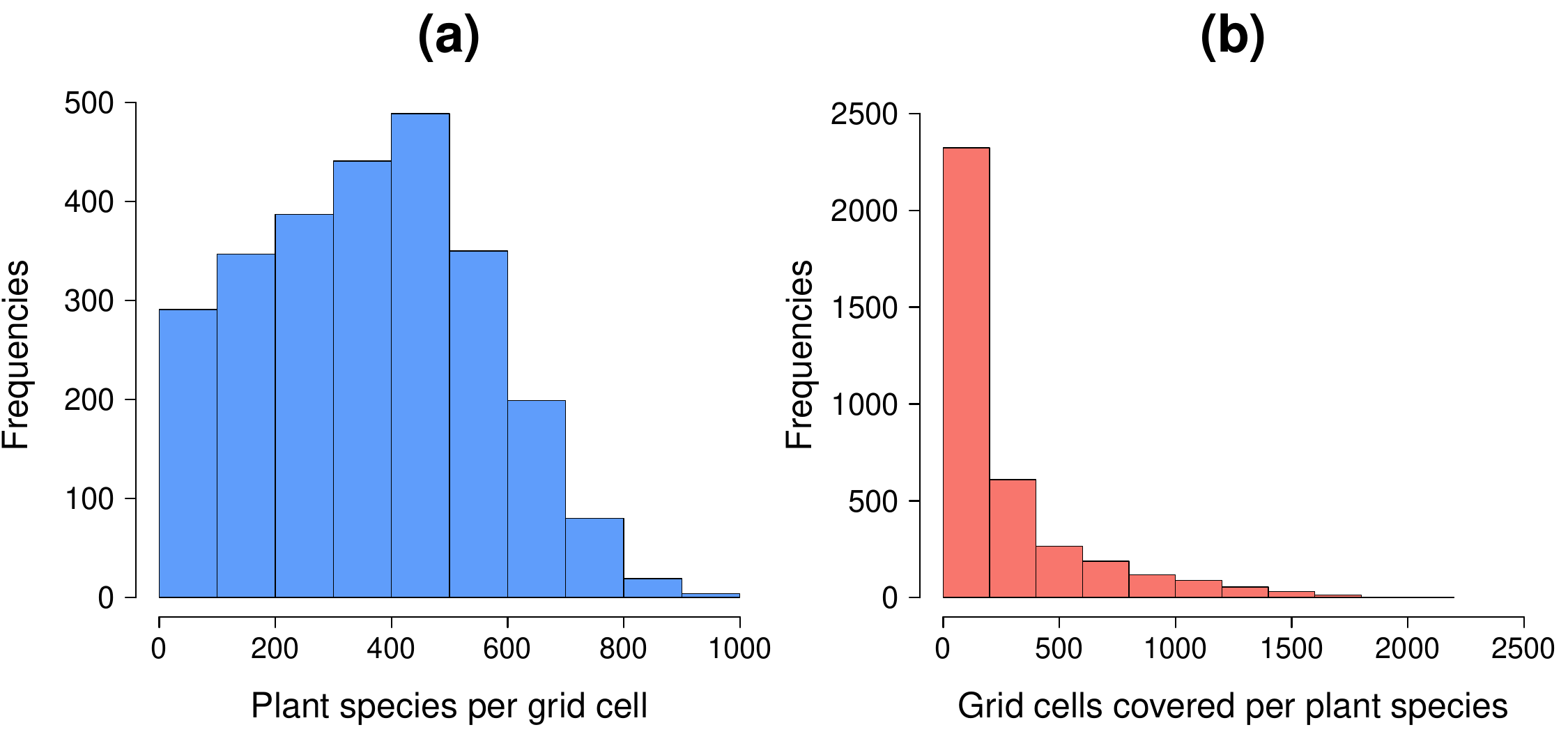}
		\caption{\sf \textbf{Histograms of the degree distributions of the biogeographical bipartite network.} Histogram of the number of plant species per grid cell (a) and the number of cells covered per plant species (b).\label{FigS12}}
	\end{center}
\end{figure} 

\begin{figure}[!h]
	\begin{center}
		\includegraphics[width=\linewidth]{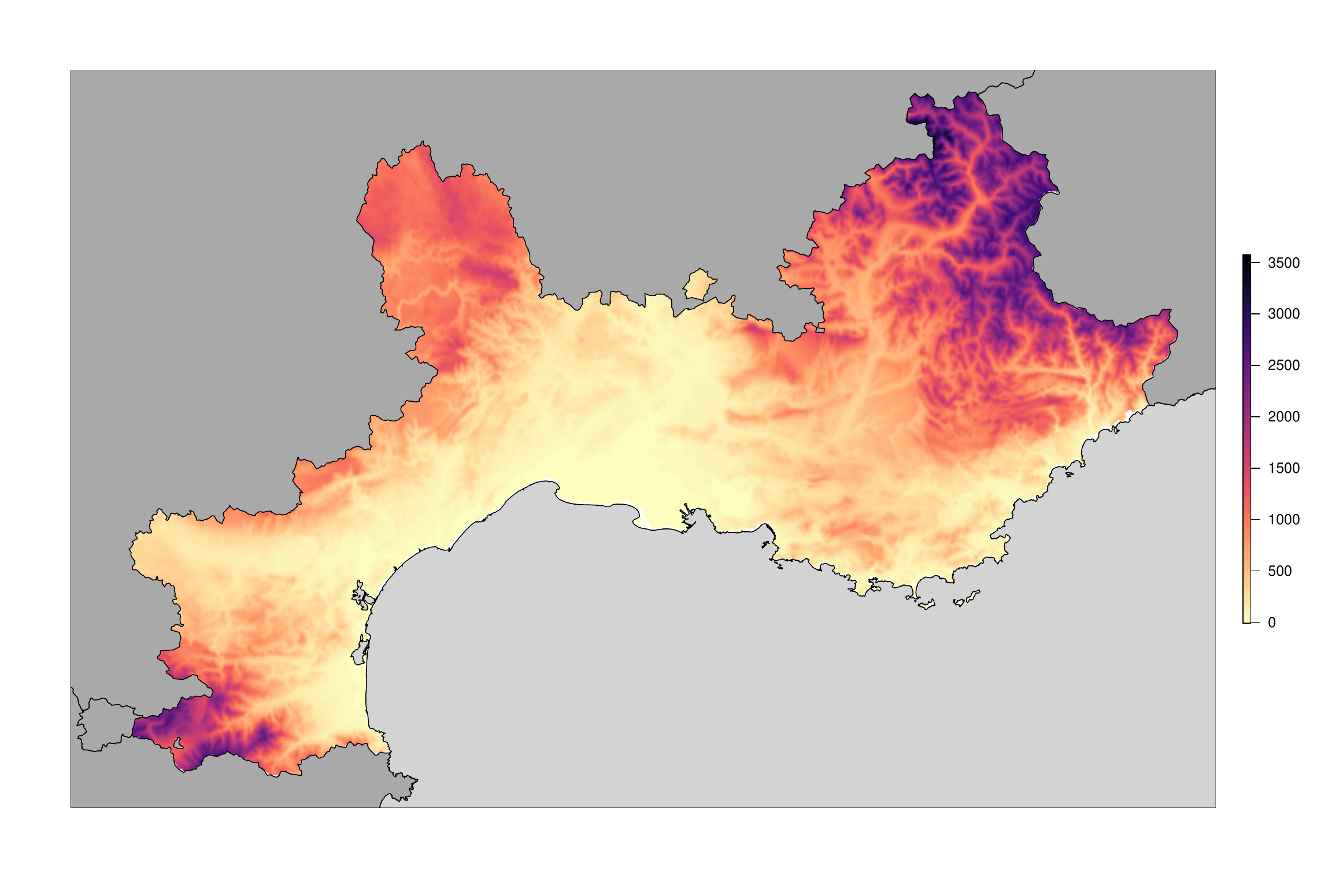}
		\caption{\sf \textbf{Altitude map of the studied area (in meters).}\label{FigS13}}
	\end{center}
\end{figure} 

\begin{figure}[!h]
	\begin{center}
		\includegraphics[width=13.5cm]{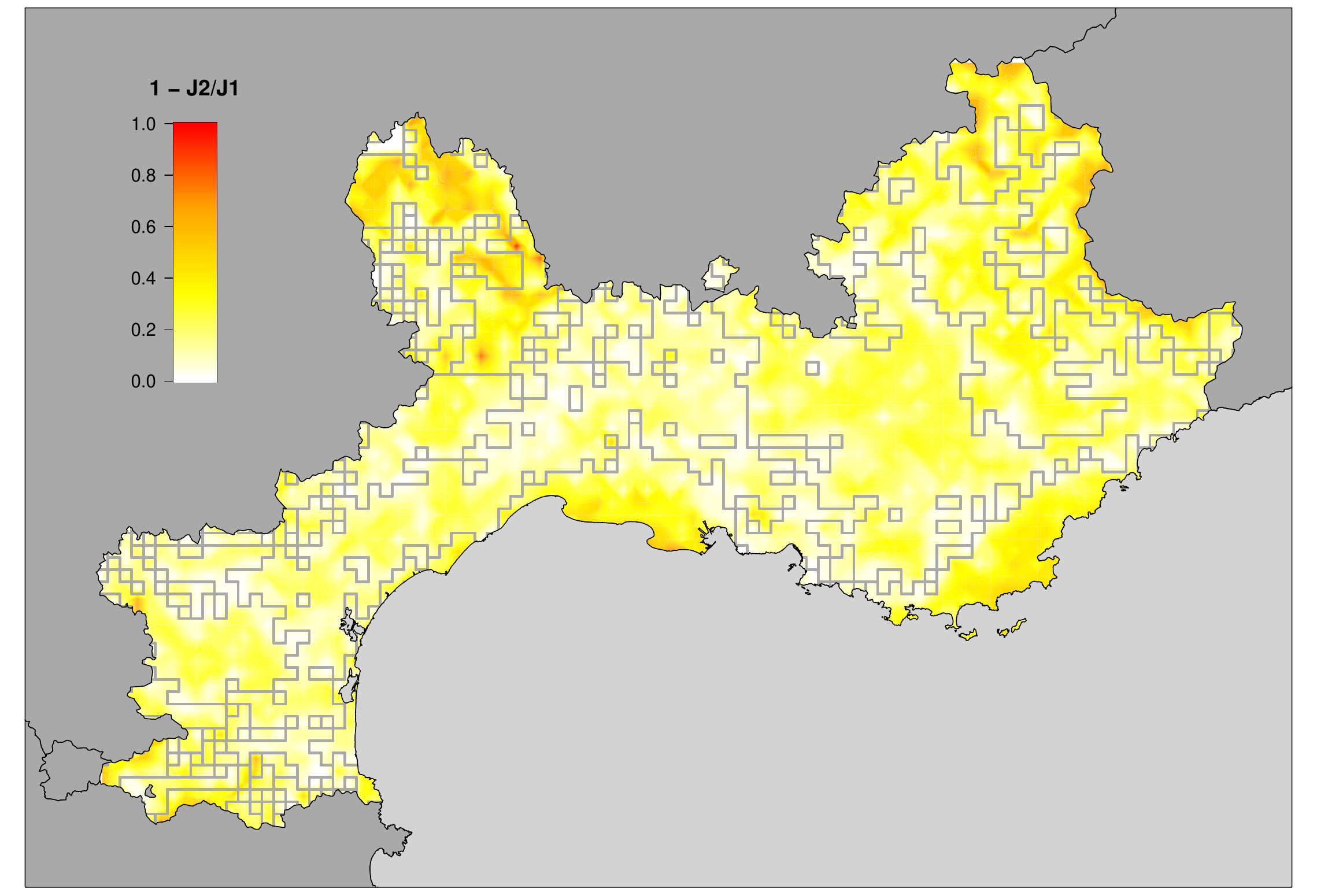}
		\caption{\sf \textbf{Uncertainty map (l = 5 km).} For a given cell, J1 represents the average Jaccard similarity index between this cell and all the cells that belong to its cluster, and J2 represents the average Jaccard similarity index between this cell and all the cells belonging to the second closest cluster (based on the Jaccard similarity). \label{FigS14}}
	\end{center}
\end{figure}

\newpage
\clearpage
\newpage
\subsection*{Supplementary Tables}

\begin{table}[!h]
	\caption{\textbf{Number of plant species per group.}}
	\label{TabS3}
	\begin{center}
		\begin{tabular}{|c|c|}			
			\hline
			\textbf{Group} & \textbf{Number of species}  \\
			\hline
			
			\textbf{a}	&	445	\\
			\hline
			\textbf{b}	&	149	\\
			\hline
			\textbf{c}	&	230	\\
			\hline
			\textbf{d}	&	299	\\
			\hline
			\textbf{e}	&	169	\\
			\hline
			\textbf{f}	&	277	\\
			\hline
			\textbf{g}	&	37	\\
			\hline
			\textbf{h}	&	180	\\
			\hline
			\textbf{i}	&	242	\\
			\hline
			\textbf{j}	&	136	\\
			\hline
			\textbf{k}	&	125	\\
			\hline
			\textbf{l}	&	95	\\
			\hline
			\textbf{m}	&	180	\\
			\hline
			\textbf{n}	&	180	\\
			\hline
			\textbf{o}	&	178	\\
			\hline
			\textbf{p}	&	186	\\
			\hline
			\textbf{q}	&	44	\\
			\hline
			\textbf{r}	&	59	\\
			\hline
			\textbf{s}	&	212	\\
			\hline
			\textbf{t}	&	274	\\
			
			\hline
			
		\end{tabular}
	\end{center}
\end{table}

\begin{table}[!h]
	\caption{\textbf{Network of interactions between biogeographical regions.}}
	\label{TabS4}
	\begin{center}
		\begin{tabular}{|c|c|c|c|c|c|c|c|c|}
			\hline
			\textbf{Bioregion} & \textbf{1} & \textbf{2} & \textbf{3} & \textbf{4} & \textbf{5} & \textbf{6} & \textbf{7} & \textbf{8} \\
			\hline
			
			\textbf{1}	&	0.52	&	0.24	&	0.15	&	0.07	&	0	&	0.01	&	0.01	&	0	\\
			\hline
			\textbf{2}	&	0.15	&	0.56	&	0.11	&	0.14	&	0.01	&	0.01	&	0.02	&	0	\\
			\hline
			\textbf{3}	&	0.13	&	0.18	&	0.42	&	0.25	&	0	&	0	&	0.01	&	0	\\
			\hline
			\textbf{4}	&	0.03	&	0.11	&	0.1	&	0.55	&	0.01	&	0.01	&	0.19	&	0.01	\\
			\hline
			\textbf{5}	&	0.01	&	0.13	&	0	&	0.07	&	0.4	&	0.24	&	0.11	&	0.05	\\
			\hline
			\textbf{6}	&	0.01	&	0.03	&	0	&	0.02	&	0.1	&	0.49	&	0.17	&	0.19	\\
			\hline
			\textbf{7}	&	0	&	0.01	&	0	&	0.16	&	0.01	&	0.04	&	0.52	&	0.26	\\
			\hline
			\textbf{8}	&	0	&	0	&	0	&	0.01	&	0.01	&	0.05	&	0.28	&	0.65	\\
			\hline

		\end{tabular}
	\end{center}
\end{table}
\end{document}